\documentclass[submission, Phys]{SciPost}

\hypersetup{
    colorlinks,
    linkcolor={red!50!black},
    citecolor={blue!50!black},
    urlcolor={blue!80!black}
}

\usepackage[bitstream-charter]{mathdesign}
\DeclareSymbolFont{usualmathcal}{OMS}{cmsy}{m}{n}
\DeclareSymbolFontAlphabet{\mathcal}{usualmathcal}

\usepackage{bm}

\DeclareMathOperator{\Tr}{Tr}

\newcommand{\cA}{\mathcal{A}}
\newcommand{\cF}{\mathcal{F}}
\newcommand{\cP}{\mathcal{P}}
\newcommand{\SU}{\mathrm{SU}}
\newcommand{\Uone}{\mathrm{U}(1)}
\newcommand{\Gint}{\mathcal{G}_{\rm int}}

\begin{document}

\begin{center}{\Large \textbf{
Wilson Holonomy and Spectral Monodromy in Spin--Orbit Rings:\\
Effective Gauge Connections and Loop Observables
}}\end{center}

\begin{center}
Nelson Bol\'ivar\textsuperscript{1,2$\star$}
\end{center}

\begin{center}
{\bf 1} Escuela de F\'isica, Universidad Central de Venezuela, C\'odigo Postal 1050, Caracas, Venezuela
\\
{\bf 2} Astrum Drive Technologies, 5850 Dallas Pkwy Unit 120B, Frisco, TX 75034, USA
\\
${}^\star$ {\small \sf nelson.e.bolivar@ucv.ve}
\end{center}

\begin{center}
\today
\end{center}

\section*{Abstract}
{\bf
A spin--orbit Hamiltonian with an effective gauge structure carries two distinct loop objects
that are routinely conflated: an energy-independent \emph{Wilson holonomy}, which organizes
interference and internal spin transport, and an energy-dependent \emph{monodromy}, which
quantizes the spectrum. We show that cleanly separating these objects supplies a precise,
computable bridge between the loop/holonomy representation of gauge theories and condensed-matter
spin--orbit transport. The construction maps a spin--orbit Hamiltonian to an effective
$\Uone$ plus internal non-Abelian connection, reduces it to a first-order transport problem, and
reads physical predictions from holonomy, monodromy, curvature, and eigenphase data. Two rings
make the separation explicit. For a Dirac (graphene) ring with Rashba coupling and Aharonov--Bohm
flux, the total holonomy factorizes exactly into a commuting $\Uone$ flux phase times an internal
spin/pseudospin holonomy, and the spectrum follows from a holonomy-eigenvalue condition. For a
Rashba--Dresselhaus ring, the internal $SU(2)$ transport is genuinely non-Abelian away from the
$\alpha=\pm\beta$ pure-gauge locus, where curvature controls path ordering; spectral quantization
then requires an explicit first-order reduction obtained by phase-space doubling of the
second-order Schr\"odinger problem. A non-Abelian Stokes formulation and Magnus expansion serve as
ordering diagnostics rather than spectral tools. Spin-network ideas enter only as historical
geometric motivation, not as a dynamical import into spintronics.
}

\vspace{10pt}
\noindent\rule{\textwidth}{1pt}
\tableofcontents\thispagestyle{fancy}
\noindent\rule{\textwidth}{1pt}
\vspace{10pt}


	\section{Introduction}
	\label{sec:intro}
	
	\subsection{Motivation: from gauge geometry to spin transport}
	Gauge theories often become most transparent when reformulated in terms of nonlocal geometric data.
	For a one-form Abelian connection, the natural variables are line holonomies and Wilson loops.
	For a two-form Abelian gauge field, such as the Kalb--Ramond field, the corresponding variables are
	surface holonomies and boundary data. These reformulations do not merely replace one notation by another:
	they expose global information, consistency conditions, and topological sectors that may be less visible
	in a purely local description. \cite{Gambini1994,GambiniPullinBook,KalbRamond1974}
	
	The present work asks whether an analogous geometric representation can be constructed for spin systems
	whose Hamiltonians admit an effective gauge reconstruction. Spin--orbit-coupled rings provide a controlled
	testing ground. Their electromagnetic Aharonov--Bohm sector is Abelian, while Rashba and Dresselhaus
	couplings generate internal non-Abelian transport. The resulting problem is therefore simple enough to
	calculate explicitly, but rich enough to exhibit path ordering, curvature, monodromy, and nontrivial loop
	observables. \cite{FroehlichStuder1993,TokatlySherman2009,Hatano2007,Recher2007}
	
	\subsection{From ``SOC as a connection'' to loop-observable physics}
	Once a connection is identified, the natural geometric object is not the local gauge potential itself but
	the \emph{parallel transport} it generates along paths. Given a $G$-connection $\cA$ on a manifold $M$, the
	holonomy (Wilson line) associated with a path $\gamma$ is
	\begin{equation}
	U(\gamma)=\cP\exp\!\left(-i\int_\gamma \cA\right)\in G,
	\end{equation}
	and for a loop $C$ the gauge-invariant Wilson loop is $W(C)=\Tr\,U(C)$.
	Holonomies encode the multiplicative composition law of paths and form the backbone of loop/holonomy
	formulations of gauge theories (including the $C^\ast$-algebraic approach to the holonomy algebra and the
	loop transform). \cite{AshtekarIsham1992,GambiniPullinBook}
	Moreover, Wilson-loop data can be used (under suitable assumptions) to reconstruct gauge potentials up to
	gauge transformations, emphasizing that holonomies are not merely secondary observables but a complete
	geometric encoding. \cite{Giles1981}
	
	In condensed-matter SOC problems, the relevant internal non-Abelian structure is typically \emph{effective}
	(determined by material parameters and external controls rather than a dynamical gauge field). The useful result is
	therefore not merely the identification of a connection, but the reconstruction of the observable loop data it controls.
	Holonomy organizes interference and spin transport on multiply connected spaces. Spectral claims require an additional
	step: the loop observable must be the monodromy of an explicitly energy-dependent first-order transport problem. This
	distinction is automatic in first-order Dirac systems, but must be constructed carefully in second-order
	Schr\"odinger-type ring Hamiltonians.

The lineage is deliberately conservative. In Abelian gauge theory, Maxwell-type one-form connections
are efficiently encoded by line holonomies and Wilson-loop variables; in higher-form Abelian systems
such as Kalb--Ramond theory, the natural nonlocal variables are surface holonomies and their boundary
data.~\cite{Gambini1994,GambiniPullinBook,KalbRamond1974} The present work extends this
geometric-representation logic to reconstructed spin--orbit gauge structures: a Hamiltonian supplies
an effective $\Uone\times\Gint$ connection, the connection supplies ordered transport, and the global
physical content is read from holonomy, monodromy, curvature, eigenphases, and loop/surface
observables.

\subsection{The central result: Hamiltonian-to-loop reconstruction}
	This work builds a structure-preserving bridge between SOC quantum matter and geometric representation variables by
	implementing a layered reconstruction:
	\begin{equation}
	\begin{gathered}
	H_{\rm SOC}
	\xrightarrow{\;\mathfrak{C}_1\;}
	\cA_{\Uone}\oplus\cA_{\rm int}
	\xrightarrow{\;\mathfrak{T}\;}
	T(E;\gamma)\\
	\xrightarrow{\;\mathfrak{H}\;}
	U(E;\gamma),\,W(C).
	\end{gathered}
	\label{eq:intro_two_comp}
	\end{equation}
	Here $\mathfrak{C}_1$ is the (model-dependent) identification of an effective $G$-connection from the Hamiltonian,
	$\mathfrak{T}$ denotes the reduction to a first-order transport problem, and $\mathfrak{H}$ maps the resulting transport generator
	to ordered exponentials and loop data. The last step is \emph{algebra preserving} in the precise sense that path concatenation is mapped to group
	multiplication and path reversal to inversion:
	\begin{equation}
	\begin{aligned}
	&U(\gamma_2\circ\gamma_1)=U(\gamma_2)\,U(\gamma_1),
	\quad
	U(\gamma^{-1})=U(\gamma)^{-1},\\
	&U(\gamma)^\dagger=U(\gamma^{-1}).
	\end{aligned}
	\label{eq:intro_algebra}
	\end{equation}
	This property is the mathematical core of the holonomy layer: it carries the loop composition algebra
	into a $\ast$-algebra of unitary operators acting on the internal Hilbert space. The additional transport layer is what
	makes the framework suitable for both interferometric phases and honest spectral quantization. The observable layer then
	translates the resulting eigenphases and monodromy conditions into flux shifts, spin-resolved interference, and persistent-current
	responses.
	
	\subsection{What is new in this construction?}
The individual ingredients are familiar in different communities: spin--orbit couplings may be
rewritten as effective non-Abelian gauge fields, Wilson lines are standard objects in gauge theory,
and ring spectra are organized by Aharonov--Bohm and Aharonov--Casher phases. Three things are new
in the way we combine them.

First, we distinguish sharply between two loop objects that are routinely conflated: the
energy-independent \emph{Wilson holonomy}, which organizes interference and internal spin transport,
and the energy-dependent \emph{monodromy}, which is the object that enters spectral quantization.
Keeping the two separate is what lets interferometric and spectral statements be made within one
framework without contradiction; conflating them is the source of recurring confusion in geometric
accounts of spin--orbit phases.

Second, this distinction supplies a concrete and \emph{synergetic link between two research areas}
that rarely meet at the level of calculation: the loop/holonomy representation of gauge theories
(Abelian path and surface variables, Wilson-loop algebras) on one side, and condensed-matter
spin--orbit transport on the other. The bridge is operational rather than analogical---each layer
of the construction maps to an explicit object that can be computed for a given ring.

Third, the starting point is not a fundamental Yang--Mills theory but a spin--orbit Hamiltonian
whose effective connection is reconstructed from material and geometric data. The result is
therefore neither an ordinary band-structure calculation nor a literal spin-network quantization,
but a transport-geometric reformulation of spin systems whose dynamics can be encoded by effective
connections.

\subsection{Main results and roadmap}
	We implement \eqref{eq:intro_two_comp} in two canonical settings and use them to separate what is kinematically geometric
	from what genuinely enters through energy-dependent transport:
	
	\paragraph{(i) Dirac (graphene) ring with Rashba SOC and AB flux.}
	For a narrow graphene ring, the AB flux enters as a central $\Uone$ holonomy, while SOC contributes an internal
	(non-Abelian) factor. This yields an exact holonomy factorization at the level of the loop observable:
	\begin{equation}
	U_{\rm tot}(\Phi)=\exp\!\left(i2\pi\frac{\Phi}{\Phi_0}\right)\,U_{\rm int},
	\label{eq:intro_factor_graphene}
	\end{equation}
	and the spectrum follows from a monodromy (holonomy-eigenvalue) condition. In this way, the familiar AB shift
	appears as a commuting $\Uone$ phase multiplying an internal spin/pseudospin holonomy, providing a geometric
	unification of AB and SOC-induced phases in ring interferometry. \cite{Recher2007,Hatano2007}
	
	\paragraph{(ii) Rashba--Dresselhaus ring as a genuinely non-Abelian transport problem.}
	For a 2DEG ring with simultaneous Rashba and Dresselhaus couplings, the effective $SU(2)$ connection
	generically has nonzero commutator curvature. A distinguished checkpoint arises on the pure-gauge locus
	(where the $SU(2)$ curvature vanishes), in which case internal holonomy becomes trivial up to conjugation.
	Away from this locus, path ordering is essential and the loop observable is controlled by commutator/curvature
	data. The ordering corrections are organized systematically by the Magnus expansion. For interferometric questions this
	already yields gauge-invariant phase information; for spectral quantization one must additionally derive an explicit
	energy-dependent first-order transport operator rather than reading the spectrum directly from a purely geometric Wilson loop.
	\cite{TokatlySherman2009,Hatano2007,Blanes2009PhysRep}
	
	\paragraph{Loop--surface lift and geometric visualization.}
	To connect loop holonomy to curvature pictures on spanning surfaces, we use a non-Abelian Stokes theorem
	and fix explicit surface-ordering conventions. This provides a uniform diagrammatic language:
	holonomy lives on loops, curvature lives on surfaces, and non-Abelianity appears as ordering/commutator
	structure. In the unified framework adopted here, this surface language is interpretive rather than a substitute
	for the underlying 1D transport calculation. \cite{Broda2000,KarpMansouriRno1999}
	
	\paragraph{Observable layer.}
	The final step is not the construction of the Wilson loop itself, but the extraction of measurable quantities from it.
	The same monodromy data that organize the spectrum also encode effective flux shifts, Aharonov--Casher phase splittings,
	spin-resolved interference, and persistent-current response. This observable-first reading retains the geometric language
	that makes the construction portable across SOC platforms.
	
	\paragraph{Relation to path/surface representations.}
	Our construction is intentionally compatible with geometric representations of gauge theories, where Abelian one-form
	fields are encoded by line holonomies and higher-form fields by surface variables and boundary data. This is the
	historical sense in which loop and spin-network ideas inform the present work. We emphasize, however, that in the
	condensed-matter setting $\cA$ is an effective/background connection reconstructed from the Hamiltonian rather than a
	dynamical Yang--Mills field. The loop upgrade is used as a
	structural and computational tool for transport and interferometry.
	
	\subsection{Organization of the paper}
	Section~\ref{sec:covariant_structure} sets up the unified $\Uone$ plus internal non-Abelian covariant language,
	introduces the effective connections in both the Pauli/2DEG and Dirac/graphene sectors, and identifies
	the pure-gauge locus $\alpha=\pm\beta$ as a structural checkpoint.
	Section~\ref{sec:loop_algebra} establishes the holonomy map and the $\ast$-algebra preservation properties
	that define the algebraic holonomy layer.
	Section~\ref{sec:dirac_ring} constructs the effective connection for a graphene Dirac ring with Rashba SOC
	and AB flux and derives the spectrum as a holonomy quantization condition.
	Section~\ref{sec:RD_ring} treats the Rashba--Dresselhaus ring, emphasizing the role of non-Abelian curvature
	and path ordering.
	Section~\ref{sec:geometry_diagrams} develops the geometric/diagrammatic construction and the loop--surface lift.
	Appendix~\ref{app:NAST} fixes the non-Abelian Stokes theorem conventions and surface ordering, while
	Appendix~\ref{app:Magnus} provides the detailed Magnus-expansion control of ordering effects and connects it
	directly to curvature in the Rashba--Dresselhaus setting.
	Additional appendices provide explicit diagonalizations and further links to loop/surface representations.

	\section{Unified \texorpdfstring{$U(1)$}{U(1)} Plus Internal Non-Abelian Covariant Structure}
	\label{sec:covariant_structure}
	
	\subsection{Internal space, structure group, and conventions}
	Our constructions use a common gauge-geometric language for two classes of systems:
	(i) nonrelativistic (Pauli/Schr\"odinger) electrons in 2DEGs with Rashba/Dresselhaus SOC; and
	(ii) Dirac (graphene) carriers with intrinsic and Rashba SOC in ring geometries.
	In both cases, the relevant structure group is taken to be
	\begin{equation}
	G=\Uone\times \Gint,
	\end{equation}
	where $\Uone$ encodes electromagnetic phases and $\Gint$ denotes the internal non-Abelian transport sector.
	For the Pauli/2DEG problem one has $\Gint=\SU(2)$ acting on spin, recovering the standard local
	$U(1)\times SU(2)$ structure emphasized by Fr\"ohlich and Studer. \cite{FroehlichStuder1993}
	For graphene, by contrast, the effective transport acts on pseudospin $\otimes$ spin, so $\Gint$ is the
	non-Abelian subgroup generated by the corresponding internal operators rather than a single fundamental-spin
	$\SU(2)$ acting on $\mathbb{C}^2$. In both settings the internal connection is \emph{effective}: it is a
	surrogate encoding of physical electric fields, crystal fields, and material parameters, rather than a
	dynamical Yang--Mills field. \cite{Hatano2007,BercheBolivarLopezMedina2015}
	
	We denote the internal Hilbert space by $\mathcal{H}_{\rm int}$.
	For the 2DEG/Pauli setting $\mathcal{H}_{\rm int}\cong\mathbb{C}^2$ (spin), while for graphene
	$\mathcal{H}_{\rm int}$ includes sublattice pseudospin (and optionally valley) tensored with real spin.
	In the Pauli sector we use generators $T^a=\sigma^a/2$ ($a=1,2,3$) with
	$[T^a,T^b]=i\epsilon^{abc}T^c$; in the graphene sector the effective internal algebra is represented directly on
	$\mathcal{H}_{\rm int}$ by the pseudospin--spin matrices appearing in the ring Hamiltonian.
	
	\subsection{Covariant derivatives and field strengths}
	Let $A_\mu$ be the electromagnetic $\Uone$ gauge field and let $W_\mu$ be a Hermitian connection one-form
	valued in the Lie algebra of $\Gint$ and acting on $\mathcal{H}_{\rm int}$. We define the $G$-covariant derivative
	\begin{equation}
	D_\mu \equiv \partial_\mu + i\frac{e}{\hbar}A_\mu\,\mathbb{I} - i\,W_\mu,
	\label{eq:Dmu_def}
	\end{equation}
	with curvature (field strength)
	\begin{equation}
	[D_\mu,D_\nu]
	=
	i\frac{e}{\hbar}F_{\mu\nu}\,\mathbb{I} - i\,\mathcal{F}_{\mu\nu},
	\qquad
	F_{\mu\nu}=\partial_\mu A_\nu-\partial_\nu A_\mu,
	\label{eq:curv_u1_su2}
	\end{equation}
	\begin{equation}
	\mathcal{F}_{\mu\nu}
	=
	\partial_\mu W_\nu-\partial_\nu W_\mu - i[W_\mu,W_\nu].
	\label{eq:curv_su2}
	\end{equation}
	Under $\Uone$ transformations $A_\mu\mapsto A_\mu-\partial_\mu\chi$ and under local internal rotations
	$g(x)\in \Gint$,
	\begin{equation}
	W_\mu \mapsto W_\mu^g = gW_\mu g^{-1} + i(\partial_\mu g)g^{-1},
	\qquad
	\mathcal{F}_{\mu\nu}\mapsto g\,\mathcal{F}_{\mu\nu}\,g^{-1}.
	\label{eq:gauge_transf_secII}
	\end{equation}
	These are the standard non-Abelian gauge transformation laws used in SOC-as-gauge formulations. \cite{TokatlySherman2009,Hatano2007}
	
	\subsection{\texorpdfstring{Pauli/2DEG sector: SOC as an $\SU(2)$ connection}{Pauli/2DEG sector: SOC as an SU(2) connection}}
	For a nonrelativistic electron in the plane with SOC, a convenient starting point is a Pauli-type Hamiltonian
	written in a minimal-coupling form with an effective $\SU(2)$ vector potential,
	\begin{equation}
	\begin{split}
	H
	&=\frac{1}{2m}\big(-i\hbar\nabla - e\bm A\,\mathbb{I} + \hbar\bm W\big)^2\\
	&\quad+V(\bm r) + \Phi^a(\bm r)\,T^a
	-\frac{\hbar^2}{2m}\bm W^{\,2}.
	\end{split}
	\label{eq:Pauli_SU2_minimal}
	\end{equation}
	Here $\bm W=(W_x,W_y)$ encodes linear-in-momentum SOC, $\Phi^aT^a$ collects Zeeman/exchange fields when present,
	and the $-\hbar^2\bm W^2/(2m)$ term is the non-Abelian analogue of a ``diamagnetic'' contribution.
	This is the standard gauge-field viewpoint in which Rashba and Dresselhaus SOC can be regarded as a
	Yang--Mills (non-Abelian) gauge field acting on spin. \cite{Hatano2007}
	
	For uniform linear Rashba ($\alpha$) and Dresselhaus ($\beta$) couplings,
	\begin{equation}
	H_{\rm SO}=
	\frac{\alpha}{\hbar}(\sigma_x p_y-\sigma_y p_x)+\frac{\beta}{\hbar}(\sigma_x p_x-\sigma_y p_y),
	\end{equation}
	one choice of constant $\bm W$ reproducing $H_{\rm SO}$ via the cross term in \eqref{eq:Pauli_SU2_minimal} is
	\begin{equation}
	W_x = \frac{m}{\hbar^2}\big(\alpha\,\sigma_y-\beta\,\sigma_x\big),
	\qquad
	W_y = \frac{m}{\hbar^2}\big(\beta\,\sigma_y-\alpha\,\sigma_x\big),
	\label{eq:W_RD_choice}
	\end{equation}
	up to an overall sign convention in \eqref{eq:Dmu_def}. This mapping is used explicitly in non-Abelian ring
	interferometry analyses and in ``SU(2) gauge'' treatments of spin filtering. \cite{Hatano2007}
	
	\paragraph{Pure-gauge checkpoint.}
	For uniform $\alpha,\beta$, the $\SU(2)$ curvature is commutator-generated,
	$\mathcal{F}_{xy}=-i[W_x,W_y]$, and vanishes on the locus $\alpha=\pm\beta$.
	Tokatly and Sherman emphasize the physical consequences of the \emph{pure-gauge} case: the SOC can be removed
	by a local $\SU(2)$ rotation, leading to strong constraints on equilibrium spin currents and to giant
	anisotropies in spin relaxation. \cite{TokatlySherman2009}
	
	\subsection{Dirac/graphene sector: effective non-Abelian potentials and limited gauge freedom}
	Near the Dirac points, graphene with spin-dependent interactions admits a formulation in terms of effective
	non-Abelian gauge potentials acting on the internal (pseudospin $\otimes$ spin) space.
	Berche \emph{et al.} show explicitly that these non-Abelian potentials are surrogates of physical fields and
	material parameters and therefore only enjoy a \emph{limited} gauge freedom: generic ``gauge transformations''
	correspond to changes of the physical model rather than redundancies. \cite{BercheBolivarLopezMedina2015}
	This perspective provides the correct conceptual bridge between fundamental gauge invariance and the effective
	$\SU(2)$ structures used in SOC quantum matter.
	
	In ring geometries, an additional geometric ingredient enters: when the Dirac operator is expressed in a
	curvilinear (polar) frame, Hermiticity requires a connection term associated with the rotating local basis.
	This term can be interpreted as a (pseudo)spin connection on $S^1$, and it is responsible for the familiar
	Berry-phase structure of half-integer angular momentum quantization in Dirac rings. (In our explicit graphene
	ring construction this is the origin of the $\sigma_\rho$ ``connection'' term that is removed by the comoving
	frame rotation in Sec.~\ref{sec:dirac_ring}.)
	
	\subsection{What is ``gauge'' in the present setting?}
	Equations \eqref{eq:Dmu_def}--\eqref{eq:gauge_transf_secII} are written in the language of gauge theory, but
	their interpretation depends on context:
	\begin{itemize}
	\item In fundamental nonrelativistic gauge formulations, local $U(1)\times SU(2)$ invariance is a structural
	principle (with $SU(2)$ acting on spin). \cite{FroehlichStuder1993}
	\item In SOC quantum matter, the $\SU(2)$ potential is typically \emph{effective} and encodes microscopic
	spin-dependent interactions; thus only specific local rotations correspond to symmetries, while generic
	transformations implement maps between physically distinct SOC configurations. \cite{BercheBolivarLopezMedina2015,TokatlySherman2009}
	\end{itemize}
	This distinction is crucial for the loop upgrade developed in Sec.~\ref{sec:loop_algebra}: our ``second composition''
	uses holonomies of the effective connection as the organizing geometric data, without assuming that $\SU(2)$ is a
	redundancy in the condensed-matter sense.
	
	\subsection{From covariant structure to loop observables}
	With the $G$-connection specified (either in the Pauli/2DEG sector via $\bm W$ and $\bm A$, or in the Dirac/graphene
	sector via the corresponding internal connection and geometric terms), the holonomy map of
	Sec.~\ref{sec:loop_algebra} assigns to each path $\gamma$ a group element
	$U(\gamma)=\cP\exp(-i\int_\gamma \cA)\in G$. This is the input for our loop/holonomy formulation of spectra and
	interference in Sec.~\ref{sec:dirac_ring} and Sec.~\ref{sec:RD_ring}, and for the surface lift in
	Sec.~\ref{sec:geometry_diagrams} and Appendix~\ref{app:NAST}.
	
	\section{Loop composition and algebra preservation}
	\label{sec:loop_algebra}
	
	\subsection{Paths, loops, and the second composition}
	Let $M$ be the configuration manifold for the effective single-particle dynamics (e.g., $M=S^1$ for a ring),
	and let $G$ be the relevant structure group. In the applications below we take
	\begin{equation}
	G=\Uone\times \Gint,
	\end{equation}
	where $\Uone$ encodes electromagnetic (AB) phases and $\Gint$ encodes the internal non-Abelian transport
	induced by spin--orbit couplings.
	
	Denote by $\Pi_1(M)$ the \emph{path groupoid} of $M$: objects are points $x\in M$, morphisms are piecewise smooth
	paths $\gamma:[0,1]\to M$ with fixed endpoints $\gamma(0)=x_i$, $\gamma(1)=x_f$, modulo reparametrizations.
	Composition is concatenation $\gamma_2\circ\gamma_1$ whenever $\gamma_1(1)=\gamma_2(0)$, and inversion is the reversed path
	$\gamma^{-1}(s)=\gamma(1-s)$.
	
	The \emph{second composition} is the assignment
	\begin{equation}
	\mathfrak{H}_\cA:\Pi_1(M)\longrightarrow G
	\end{equation}
	defined by parallel transport of an effective connection $\cA$ (defined below). This map upgrades
	the original quantum system into a loop-/path-based representation where the primary objects are holonomies
	(or Wilson lines/loops) rather than local gauge potentials.
	
	\subsection{Effective connection and parallel transport}
	Let $\cA$ be a Lie-algebra valued one-form on $M$ in a fixed unitary representation on the internal Hilbert space
	$\mathcal{H}_{\rm int}$ (spin, pseudospin, valley, etc.):
	\begin{equation}
	\cA = \cA_\mu(x)\,dx^\mu, \qquad \cA_\mu(x)^\dagger=\cA_\mu(x)
	\end{equation}
	so that $-i\,\cA_\mu$ is anti-Hermitian and generates unitary transport.  Along a path $\gamma(s)$ we define
	the evolution operator $U_\gamma(s)$ by the transport equation
	\begin{equation}
	\frac{d}{ds}U_\gamma(s)= -\,i\,\dot{\gamma}^\mu(s)\,\cA_\mu(\gamma(s))\,U_\gamma(s),
	\qquad
	U_\gamma(0)=\mathbb{I},
	\label{eq:transport_eq}
	\end{equation}
	and set the \emph{Wilson line / holonomy along $\gamma$} to be
	\begin{equation}
	U_\gamma[\cA]\equiv U_\gamma(1)=\mathcal{P}\exp\!\left(-i\int_\gamma \cA\right)\in G.
	\label{eq:holonomy_def}
	\end{equation}
	For a closed loop $\gamma$ based at $x$ (i.e.\ $\gamma(0)=\gamma(1)=x$), $U_\gamma[\cA]$ is the holonomy at $x$.
	
	\subsection{Algebraic properties: multiplicativity, inversion, and \texorpdfstring{$*$}{*}-structure}
	The following properties implement the desired ``algebra preservation'' of the second composition.
	
	\paragraph{Lemma 1 (Reparametrization invariance).}
	If $\gamma$ and $\tilde\gamma$ differ only by an orientation-preserving reparametrization,
	then $U_\gamma[\cA]=U_{\tilde\gamma}[\cA]$.
	
	\paragraph{Lemma 2 (Concatenation).}
	Let $\gamma_1:x_0\to x_1$ and $\gamma_2:x_1\to x_2$. Then
	\begin{equation}
	U_{\gamma_2\circ\gamma_1}[\cA]=U_{\gamma_2}[\cA]\;U_{\gamma_1}[\cA].
	\label{eq:concat}
	\end{equation}
	
	\paragraph{Lemma 3 (Inverse path).}
	For $\gamma:x_0\to x_1$,
	\begin{equation}
	U_{\gamma^{-1}}[\cA]=U_\gamma[\cA]^{-1}.
	\label{eq:inverse}
	\end{equation}
	
	\paragraph{Lemma 4 (Unitarity and \texorpdfstring{$*$}{*}-structure).}
	If $\cA_\mu^\dagger=\cA_\mu$ (equivalently $-i\cA$ anti-Hermitian), then
	\begin{equation}
	U_\gamma[\cA]^\dagger=U_\gamma[\cA]^{-1}=U_{\gamma^{-1}}[\cA].
	\label{eq:star}
	\end{equation}
	
	\noindent
	\emph{Proof sketch (Lemmas 2--4).}
	Equation \eqref{eq:transport_eq} implies uniqueness of solutions; concatenation corresponds to solving
	\eqref{eq:transport_eq} on $[0,1/2]$ and $[1/2,1]$ and matching at the midpoint, yielding \eqref{eq:concat}.
	The inverse-path statement follows by changing variables $s\mapsto 1-s$ and using $U(0)=\mathbb{I}$.
	Finally, differentiating $U_\gamma^\dagger U_\gamma$ and using $\cA_\mu^\dagger=\cA_\mu$ shows $\frac{d}{ds}(U^\dagger U)=0$,
	so $U^\dagger U=\mathbb{I}$ and \eqref{eq:star} follows.
	
	\subsection{Gauge covariance and gauge-invariant loop observables}
	Under a (local) $G$-valued transformation $g:M\to G$, the connection transforms as
	\begin{equation}
	\cA \mapsto \cA^g = g\,\cA\,g^{-1} + i\,dg\,g^{-1},
	\label{eq:gauge_transf}
	\end{equation}
	and the Wilson line transforms covariantly at endpoints:
	\begin{equation}
	U_\gamma[\cA^g]=g(x_f)\,U_\gamma[\cA]\,g(x_i)^{-1}.
	\label{eq:holonomy_gauge_cov}
	\end{equation}
	Therefore, for loops $\gamma$ based at $x$ the conjugacy class of $U_\gamma$ is gauge invariant, and the traced holonomy
	in a unitary representation $R$,
	\begin{equation}
	W_\gamma^{(R)}[\cA]\equiv \mathrm{Tr}_R\,U_\gamma[\cA],
	\label{eq:wilson_loop_def}
	\end{equation}
	is gauge invariant. In loop-based approaches, the set $\{W_\gamma\}$ provides a natural generating set for gauge-invariant observables,
	subject to group identities (e.g.\ Mandelstam-type relations in $\SU(2)$).
	
	\subsection{Holonomy algebras and quantum states as positive functionals}
	Let $\mathfrak{A}_{\rm hol}$ be the $\ast$-algebra generated by matrix elements of $U_\gamma[\cA]$ (open paths) or by traced holonomies
	$W_\gamma[\cA]$ (closed loops), equipped with the involution induced by \eqref{eq:star}. One may complete this algebra to a
	$C^\ast$-algebra of holonomy functions on (a suitable completion of) the space of connections, providing a natural configuration algebra
	for loop-based quantization.
	
	In the present condensed-matter setting we will often work in a fixed internal Hilbert space $\mathcal{H}_{\rm int}$ and treat $\cA$ as an effective
	(background) connection determined by material parameters and external fields. In that case, any density matrix $\rho$ on $\mathcal{H}_{\rm int}$
	defines a positive linear functional (a state) on $\mathfrak{A}_{\rm hol}$ by
	\begin{align}
	\omega_\rho\!\left(U_\gamma\right)&=\mathrm{Tr}\!\left(\rho\,U_\gamma[\cA]\right),\notag\\
	\omega_\rho\!\left(W_\gamma^{(R)}\right)&=\mathrm{Tr}\!\left(\rho\,\mathrm{Tr}_R\,U_\gamma[\cA]\right),
	\label{eq:state_functional}
	\end{align}
	thereby preserving the standard quantum-mechanical $\ast$-algebraic structure while recasting the theory in loop/holonomy variables.
	
	\subsection{Surface lift (for geometric interpretation)}
	When $\gamma=\partial\Sigma$ bounds an oriented surface $\Sigma\subset M$ and the connection is sufficiently regular,
	the holonomy admits a (non-Abelian) surface representation in terms of curvature
	\begin{equation}
	\cF = d\cA - i\,\cA\wedge \cA,
	\end{equation}
	schematically written as a surface-ordered exponential (with ``twisted'' curvature)---the non-Abelian Stokes theorem.
	This provides the geometric dictionary used in our figures: \emph{holonomy lives on loops}, while \emph{curvature lives on spanning surfaces}.
	
	\section{Dirac (graphene) ring: Rashba + AB flux as a Wilson-loop quantization}
	\label{sec:dirac_ring}
	
	\subsection{Continuum model and symmetry sector}
	We consider a graphene monolayer in the low-energy Dirac regime, including spin--orbit interactions.
	In a single-valley description (valley index $\tau=\pm1$ treated as a good quantum number when intervalley
	scattering is weak), the effective Hamiltonian can be written as the Dirac kinetic term plus intrinsic and
	Rashba SOC contributions,
	\begin{equation}
	\begin{aligned}
	H_{\rm 2D}
	&=\hbar v_F\big(\tau\,\sigma_x \Pi_x + \sigma_y \Pi_y\big)
	+\Delta_{\rm SO}\,\tau\,\sigma_z s_z\\
	&\quad+\lambda_R\big(\tau\,\sigma_x s_y-\sigma_y s_x\big)
	+V_{\rm conf}(\bm r),
	\end{aligned}
	\label{eq:H2D_graphene_SOC}
	\end{equation}
	where $\bm{\sigma}$ acts on sublattice (pseudospin), $\bm{s}$ on real spin, and
	$\bm{\Pi}=\bm p+e\bm A$ is the minimally coupled momentum.
	The low-energy form \eqref{eq:H2D_graphene_SOC} and the structure of intrinsic/Rashba terms in graphene
	are standard \cite{Min2006} (see also Refs.~\cite{Schelter2012,Recher2007} for ring geometries).
	
	In what follows we focus on a single valley ($\tau$ fixed) and assume a smooth ring confinement
	$V_{\rm conf}$ such that intervalley mixing is negligible, as in analytic Dirac-ring models. \cite{Recher2007,Schelter2012}
	
	\subsection{Ring reduction and the spin connection}
	We restrict to a narrow ring of radius $a$ and introduce polar angle $\varphi\in[0,2\pi)$.
	Define the local pseudospin matrices
	\begin{equation}
	\begin{aligned}
	\sigma_\rho(\varphi)
	&=\sigma_x\cos\varphi+\sigma_y\sin\varphi,\\
	\sigma_\varphi(\varphi)
	&=-\sigma_x\sin\varphi+\sigma_y\cos\varphi,
	\end{aligned}
	\end{equation}
	and analogously $s_\rho(\varphi),s_\varphi(\varphi)$ in the real-spin space.
	The Dirac operator in curvilinear coordinates carries a spin-connection term that is required for
	Hermiticity and encodes the Berry-phase structure of spin-$1/2$ transport on a loop. \cite{BerryMondragon1987}
	
	Introducing the ring energy scale
	\begin{equation}
	\varepsilon \equiv \frac{\hbar v_F}{a},
	\end{equation}
	the effective 1D graphene-ring Hamiltonian takes the form
	\begin{equation}
	\begin{aligned}
	H(\Phi)&=
	-i\,\varepsilon\Big(\sigma_\varphi(\varphi)\,D_\varphi-\tfrac12\sigma_\rho(\varphi)\Big)\\
	&\quad+\Delta_{\rm SO}\,\sigma_z s_z
	+\lambda_R\Big(\sigma_\rho(\varphi)s_\varphi(\varphi)-\sigma_\varphi(\varphi)s_\rho(\varphi)\Big),
	\end{aligned}
	\label{eq:Hr_phi}
	\end{equation}
	where $D_\varphi$ is the AB-flux covariant derivative defined below.
	Equation \eqref{eq:Hr_phi} is the natural ring specialization of \eqref{eq:H2D_graphene_SOC} in a local frame:
	the AB sector enters as a $\Uone$ connection while Rashba generates an internal non-Abelian transport, in the
	spirit of the gauge-field formulation of SOC. \cite{Hatano2007}
	
	\subsection{AB flux as a commuting \texorpdfstring{$\Uone$}{U(1)} factor}
	A magnetic flux $\Phi$ threading the ring is implemented by the azimuthal gauge potential
	$A_\varphi=\Phi/(2\pi a)$, so that
	\begin{equation}
	D_\varphi \equiv \partial_\varphi + i\frac{\Phi}{\Phi_0},
	\qquad
	\Phi_0=\frac{h}{|e|}.
	\label{eq:Dphi}
	\end{equation}
	This is the standard Aharonov--Bohm coupling in graphene rings and produces $\Phi_0$-periodic spectra and
	interference patterns. \cite{Recher2007,Schelter2012,Wurm2009}
	
	\subsection{Comoving frame and constant-coefficient transport}
	To expose the loop/holonomy structure, it is convenient to remove the explicit $\varphi$ dependence by a
	comoving unitary rotation in pseudospin and spin spaces,
	\begin{equation}
	\begin{aligned}
	U(\varphi)
	&=\exp\!\Big(-\frac{i}{2}\varphi\,\sigma_z\Big)
	  \exp\!\Big(-\frac{i}{2}\varphi\,s_z\Big),\\
	\Psi(\varphi)&=U(\varphi)\,\chi(\varphi).
	\end{aligned}
	\label{eq:comovingU}
	\end{equation}
	Using $U^\dagger\sigma_\rho U=\sigma_x$, $U^\dagger\sigma_\varphi U=\sigma_y$ (and similarly for $s_\rho,s_\varphi$),
	together with $U^\dagger\partial_\varphi U = -\frac{i}{2}(\sigma_z+s_z)$, one obtains the constant-coefficient Hamiltonian
	\begin{align}
	H'(\Phi)\equiv U^\dagger H(\Phi)U
	&=
	-i\,\varepsilon\,\sigma_y\,D_\varphi
	-\frac{\varepsilon}{2}\sigma_y s_z
	+\Delta_{\rm SO}\,\sigma_z s_z\notag\\
	&\quad
	+\lambda_R\big(\sigma_x s_y-\sigma_y s_x\big).
	\label{eq:Hr_const}
	\end{align}
	For the Wilson-loop construction it is sufficient that \eqref{eq:Hr_const} is independent of $\varphi$.
	
	\subsection{Effective connection and Wilson loop on \texorpdfstring{$S^1$}{S1}}
	The eigenvalue problem $H'(\Phi)\chi = E\chi$ can be rewritten as a first-order parallel-transport equation
	along the loop. Define
	\begin{equation}
	V \equiv
	-\frac{\varepsilon}{2}\sigma_y s_z
	+\Delta_{\rm SO}\,\sigma_z s_z
	+\lambda_R\big(\sigma_x s_y-\sigma_y s_x\big),
	\end{equation}
	so that \eqref{eq:Hr_const} reads $-i\varepsilon\sigma_y D_\varphi\chi + V\chi=E\chi$.
	Multiplying by $\sigma_y$ and isolating $\partial_\varphi\chi$ gives
	\begin{equation}
	\begin{aligned}
	\partial_\varphi\chi
	&= i\,\mathcal{A}_\varphi(E;\Phi)\,\chi,\\
	\mathcal{A}_\varphi(E;\Phi)
	&=
	\frac{\Phi}{\Phi_0}\,\mathbb{I}
	+\frac{1}{\varepsilon}\,\sigma_y\,(E - V).
	\end{aligned}
	\label{eq:Aphi_graphene}
	\end{equation}
	This exhibits explicitly the decomposition
	\begin{equation}
	\mathcal{A}_\varphi(E;\Phi)=\underbrace{\Big(\frac{\Phi}{\Phi_0}\Big)\mathbb{I}}_{\Uone\text{ (AB) }}
	+\underbrace{\mathcal{A}^{\rm int}_\varphi(E)}_{\text{internal transport (SOC)}},
	\end{equation}
	where the AB sector is proportional to the identity and therefore commutes with the internal (spin/pseudospin)
	connection. The Wilson loop (holonomy) around the ring is
	\begin{equation}
	\begin{aligned}
	W(E;\Phi)
	&=\mathcal{P}\exp\!\left(
	i\int_0^{2\pi}\!d\varphi\;\mathcal{A}_\varphi(E;\Phi)
	\right)\\
	&=\exp\!\big(i2\pi\,\mathcal{A}_\varphi(E;\Phi)\big),
	\end{aligned}
	\label{eq:W_graphene}
	\end{equation}
	and factorizes as
	\begin{equation}
	\begin{aligned}
	&W(E;\Phi)=e^{\,i2\pi\Phi/\Phi_0}\;W_{\rm int}(E),\\
	&W_{\rm int}(E)=\exp\!\left(i\frac{2\pi}{\varepsilon}\,\sigma_y(E-V)\right).
	\end{aligned}
	\label{eq:factor_graphene}
	\end{equation}
	The unified AB (vector potential) plus spin-dependent (non-Abelian) phase structure is the
	standard gauge-field viewpoint on SOC interference. \cite{Hatano2007}
	
	\subsection{Quantization as a holonomy eigenvalue condition}
	Impose the boundary condition
	\begin{equation}
	\chi(\varphi+2\pi)=e^{-i\theta_0}\chi(\varphi),
	\label{eq:bc_theta0}
	\end{equation}
	where $\theta_0$ encodes possible geometric/twist contributions (including conventions in the comoving frame).
	Then single-valuedness is equivalent to the monodromy condition
	\begin{equation}
	\begin{aligned}
	W(E;\Phi)\,\chi(0)&=e^{-i\theta_0}\chi(0),\\
	\Longleftrightarrow\qquad
	\det\!\big(W(E;\Phi)-e^{-i\theta_0}\mathbb{I}\big)&=0.
	\end{aligned}
	\label{eq:quant_graphene}
	\end{equation}
	Using \eqref{eq:factor_graphene}, AB flux contributes only through the commuting $\Uone$ factor
	$e^{i2\pi\Phi/\Phi_0}$, i.e.\ as a shift of the total phase in the holonomy eigenvalue condition.
	
	\subsection{Closed-form spectrum for Rashba coupling}
	To make contact with explicit energy levels, we set $\Delta_{\rm SO}=0$ (Rashba-only graphene ring).
	Since \eqref{eq:Hr_const} is constant in $\varphi$, one may use eigenmodes $\chi(\varphi)=e^{im\varphi}\chi_0$
	and replace $D_\varphi\to i\tilde m$, with the effective angular quantum number
	\begin{equation}
	\tilde m \equiv m+\frac{\Phi}{\Phi_0},
	\qquad m\in\mathbb{Z}
	\end{equation}
	(including any additional twists can be absorbed into an effective shift of $m$ via $\theta_0$ in
	\eqref{eq:bc_theta0}--\eqref{eq:quant_graphene}).
	Diagonalizing the resulting $4\times4$ matrix yields the four branches
	\begin{equation}
	\begin{aligned}
	E^{\kappa,\delta}_{m}(\Phi)
	&=\frac{\kappa}{2}\Big[
	\varepsilon^2\big(1+4\tilde m^{\,2}\big)+8\lambda_R^2\\
	&\qquad
	-4\delta\sqrt{
	\big(\tilde m^{\,2}\varepsilon^2+\lambda_R^2\big)
	\big(\varepsilon^2+4\lambda_R^2\big)}
	\Big]^{1/2},
	\end{aligned}
	\label{eq:graphene_rashba_levels}
	\end{equation}
	\vspace{-3mm}
	\begin{equation*}
	\kappa=\pm1,\;\delta=\pm1.
	\end{equation*}
	Here $\kappa$ labels electron/hole branches and $\delta$ labels the SOC-split branches.
	Equation \eqref{eq:graphene_rashba_levels} is the explicit realization of the holonomy quantization
	\eqref{eq:quant_graphene} for the Rashba-only ring: AB flux enters through the universal shift
	$m\to m+\Phi/\Phi_0$, and the internal structure is controlled by the non-Abelian factor $W_{\rm int}(E)$.
	
	\paragraph{Checks.}
	(i) $\lambda_R\to0$ reproduces a Dirac-ring spectrum with $\Phi_0$-periodic AB dependence, consistent with the
	graphene-ring AB literature. \cite{Recher2007,Schelter2012,Wurm2009}
	(ii) The factorization \eqref{eq:factor_graphene} shows that AB and SOC phases combine multiplicatively at the
	level of the Wilson loop, precisely the loop-space composition that will be generalized in Sec.~\ref{sec:RD_ring}.
	
	\section{Rashba--Dresselhaus ring: non-Abelian curvature and loop observables}
	\label{sec:RD_ring}
	
	\subsection{2DEG Hamiltonian and \texorpdfstring{$U(1)\times SU(2)$}{U(1)xSU(2)} minimal coupling}
	We now turn to the nonrelativistic (Pauli/Schr\"odinger) setting, which provides the cleanest arena for
	the full Rashba--Dresselhaus (RD) non-Abelian structure. Consider a single-mode quantum ring of radius
	$a$ defined in a two-dimensional electron gas (2DEG), threaded by a magnetic flux $\Phi$ and subject to
	linear Rashba and linear Dresselhaus spin--orbit couplings. The continuum Hamiltonian is
	\begin{equation}
	H
	=
	\frac{1}{2m}\big(\bm p + e\bm A\big)^2 + V_{\rm conf}(\bm r)
	+H_{\rm SO},
	\label{eq:H_2deg}
	\end{equation}
	with
	\begin{equation}
	H_{\rm SO}
	=
	\frac{\alpha}{\hbar}\big(\sigma_x p_y-\sigma_y p_x\big)
	+\frac{\beta}{\hbar}\big(\sigma_x p_x-\sigma_y p_y\big),
	\label{eq:RD_SO}
	\end{equation}
	where $\alpha$ and $\beta$ are the Rashba and Dresselhaus strengths and $\bm\sigma$ acts on the
	spin-$\frac12$ subspace.
	
	Following the non-Abelian gauge-field viewpoint of spin--orbit coupling, one rewrites
	\eqref{eq:H_2deg}--\eqref{eq:RD_SO} as a minimal-coupling form with an effective $SU(2)$ vector potential
	$\bm{\mathcal A}$ (a matrix in spin space):
	\begin{equation}
	H
	=
	\frac{1}{2m}\big(\bm p + e\bm A - \bm{\mathcal A}\big)^2
	+V_{\rm conf}(\bm r)
	-\frac{1}{2m}\bm{\mathcal A}^{\,2}.
	\label{eq:min_coupling_SU2}
	\end{equation}
	Choosing
	\begin{equation}
	\mathcal A_x = \frac{m}{\hbar}\big(\alpha\,\sigma_y-\beta\,\sigma_x\big),\qquad
	\mathcal A_y = \frac{m}{\hbar}\big(\beta\,\sigma_y-\alpha\,\sigma_x\big),
	\label{eq:Axy_RD}
	\end{equation}
	the cross term $-(1/m)\,(p_x\mathcal A_x+p_y\mathcal A_y)$ reproduces exactly $H_{\rm SO}$ in
	\eqref{eq:RD_SO}. The remaining scalar term $-\bm{\mathcal A}^2/(2m)$ plays the role of a
	(non-Abelian) ``diamagnetic'' contribution and is central in discussions of which parts of the
	$SU(2)$ structure correspond to true gauge redundancy versus physical symmetry in condensed matter
	(we return to this point when discussing loop observables).
	
	It is convenient to define an $SU(2)$ connection in dimensionless form by
	\begin{equation}
	\bm W \equiv \frac{1}{\hbar}\bm{\mathcal A},
	\qquad
	D_i \equiv \partial_i + i\frac{e}{\hbar}A_i - i W_i,
	\label{eq:covD_RD}
	\end{equation}
	so that the kinetic term is $-(\hbar^2/2m)D_i D_i$ (up to ordering issues fixed by the chosen
	ring reduction). In this notation the structure group is $G=\Uone\times\SU(2)$.
	
	\subsection{Non-Abelian curvature and the pure-gauge locus}
	The $SU(2)$ field strength (curvature) is
	\begin{equation}
	F_{ij}
	=
	\partial_i W_j-\partial_j W_i - i[W_i,W_j].
	\label{eq:Fij_def}
	\end{equation}
	For spatially uniform $\alpha,\beta$ the derivative terms vanish and the curvature is purely
	commutator-generated. Using \eqref{eq:Axy_RD} one finds
	\begin{equation}
	F_{xy}
	=
	-\,i[W_x,W_y]
	=
	+2\left(\frac{m}{\hbar^2}\right)^2(\alpha^2-\beta^2)\,\sigma_z,
	\label{eq:Fxy_RD}
	\end{equation}
	The key point is that the
	curvature is proportional to $(\alpha^2-\beta^2)$ and therefore vanishes on the locus
	\begin{equation}
	\alpha=\pm\beta.
	\label{eq:pure_gauge_locus}
	\end{equation}
	On \eqref{eq:pure_gauge_locus} the $SU(2)$ field is \emph{pure gauge}: it can be removed by a local
	spin rotation, yielding an exact $SU(2)$ symmetry and the associated persistent-spin-helix structure.
	This locus provides a stringent internal checkpoint for our loop/holonomy formulation: when $F_{xy}=0$,
	non-Abelian path ordering becomes gauge-trivial and the spin-orbit contribution to the Wilson loop
	reduces (up to conjugation) to the identity.
	On the ring this statement should be read with the usual local/global qualification: vanishing local
	curvature removes the non-Abelian commutator content, but boundary twists, frame conventions, and winding
	data may still appear as global monodromy phases.
	
	\subsection{Ring reduction: internal Wilson loop versus spectral monodromy}
	Restrict now to a narrow ring of radius $a$ in the $xy$ plane. Let $\varphi$ be the azimuthal coordinate
	and $\hat{\bm e}_\varphi=(-\sin\varphi,\cos\varphi)$ the unit tangent. The tangential $SU(2)$ connection is
	\begin{equation}
	W_\varphi(\varphi)
	=
	a\,\hat{\bm e}_\varphi\cdot \bm W
	=
	a\big(-\sin\varphi\,W_x + \cos\varphi\,W_y\big),
	\label{eq:Wphi_def}
	\end{equation}
	which is generally \emph{$\varphi$-dependent} and points along a rotating axis in spin space when both
	$\alpha$ and $\beta$ are present.
	
	For a closed traversal of the ring the internal geometric transport associated with the effective
	$G=\Uone\times\SU(2)$ connection is
	\begin{equation}
	\begin{aligned}
	U_{\rm tot}(\Phi)
	&=
	\mathcal{P}\exp\!\left(
	i\int_0^{2\pi}\!d\varphi\;
	\left[\frac{\Phi}{\Phi_0}\,\mathbb{I}
	- W_\varphi(\varphi)\right]\right),\\
	\Phi_0&=\frac{h}{|e|}.
	\end{aligned}
	\label{eq:Utot_RD}
	\end{equation}
	Since the AB term is proportional to $\mathbb{I}$, it commutes with the internal transport and one has
	the exact factorization
	\begin{equation}
	\begin{aligned}
	&U_{\rm tot}(\Phi)=
	e^{\,i2\pi\Phi/\Phi_0}\;
	U_{\rm int},\\
	&U_{\rm int}\equiv \mathcal{P}\exp\!\left(-i\int_0^{2\pi}\!d\varphi\;W_\varphi(\varphi)\right).
	\end{aligned}
	\label{eq:factor_RD}
	\end{equation}
	The gauge-invariant loop observables introduced in Sec.~\ref{sec:loop_algebra} are then built from
	$U_{\rm tot}$, e.g.\ the Wilson loop $W(\gamma)=\mathrm{Tr}\,U_{\rm tot}$, whose phase controls
	interference in ring interferometers and underlies spin-filtering conditions. This object should not,
	however, be confused with the spectral monodromy of the Schr\"odinger problem. The latter must know about
	the eigenvalue $E$ and therefore lives naturally on a doubled phase space, as constructed next.
	
	\subsection{From the second-order ring equation to an energy-dependent transport operator}
	The loop operator \eqref{eq:factor_RD} captures the geometric spin rotation generated by the tangential
	$SU(2)$ connection, but by itself it does \emph{not} provide a spectral quantization rule because it carries
	no explicit dependence on the eigenvalue $E$. To obtain a genuine monodromy condition one must start from the
	stationary Schr\"odinger problem on the ring and convert the resulting second-order equation into a first-order
	transport system. This is the technical step that makes the RD sector comparable to the Dirac ring:
	graphene is first order from the start, whereas the nonrelativistic ring becomes first order only after
	phase-space doubling.
	
	At the single-mode level, a self-adjoint ring reduction of \eqref{eq:min_coupling_SU2} has the generic form
	\begin{equation}
	\left[-\frac{\hbar^2}{2ma^2}D_\varphi^2 + V_{\rm eff}(\varphi)\right]\psi(\varphi)=E\,\psi(\varphi),
	\label{eq:RD_ring_second_order}
	\end{equation}
	where $\psi(\varphi)\in\mathcal{H}_{\rm int}\cong\mathbb{C}^2$ is the spinor on the ring,
	$V_{\rm eff}$ includes the scalar terms produced by confinement and ordering, and
	\begin{equation}
	D_\varphi=\partial_\varphi + X(\varphi),
	\qquad
	X(\varphi)\equiv i\frac{\Phi}{\Phi_0}\,\mathbb{I}-iW_\varphi(\varphi).
	\label{eq:RD_Dphi}
	\end{equation}
	Expanding \eqref{eq:RD_ring_second_order} gives a matrix second-order ordinary differential equation
	\begin{equation}
	\psi'' + B_1(\varphi)\,\psi' + B_0(E;\varphi)\,\psi = 0,
	\label{eq:RD_second_order_expanded}
	\end{equation}
	with
	\begin{equation}
	\begin{aligned}
	B_1(\varphi)&=2X(\varphi),\\
	B_0(E;\varphi)
	&=X'(\varphi)+X(\varphi)^2
	+\frac{2ma^2}{\hbar^2}\big(E-V_{\rm eff}(\varphi)\big).
	\end{aligned}
	\label{eq:RD_B0B1}
	\end{equation}
	
	Now define the doubled state on $\mathcal{H}_{\rm int}\oplus\mathcal{H}_{\rm int}$,
	\begin{equation}
	\Psi(\varphi)\equiv
	\begin{pmatrix}
	\psi(\varphi)\\
	\psi'(\varphi)
	\end{pmatrix}.
	\label{eq:RD_doubled_state}
	\end{equation}
	Equation \eqref{eq:RD_second_order_expanded} is then equivalent to the first-order transport equation
	\begin{equation}
	\partial_\varphi \Psi(\varphi)=\mathbb{K}(E;\varphi)\,\Psi(\varphi),
	\label{eq:RD_first_order_transport}
	\end{equation}
	with the $4\times4$ transport generator
	\begin{equation}
	\mathbb{K}(E;\varphi)=
	\begin{pmatrix}
	0 & \mathbb{I}\\
	-B_0(E;\varphi) & -B_1(\varphi)
	\end{pmatrix}.
	\label{eq:RD_transport_generator}
	\end{equation}
	Equivalently, one may use a covariant doubled state
	\begin{equation}
	\Xi(\varphi)\equiv
	\begin{pmatrix}
	\psi(\varphi)\\
	\Pi_\varphi\psi(\varphi)
	\end{pmatrix},
	\qquad
	\Pi_\varphi\equiv -i\partial_\varphi+\frac{\Phi}{\Phi_0}-W_\varphi(\varphi),
	\label{eq:RD_covariant_doubled_state}
	\end{equation}
	for which the same second-order equation is written as
	\begin{equation}
	\partial_\varphi\Xi(\varphi)
	=i\,\mathbb{A}_{\rm RD}(E;\varphi)\,\Xi(\varphi).
	\label{eq:RD_covariant_transport}
	\end{equation}
	For uniform linear Rashba--Dresselhaus couplings, the scalar piece
	$W_x^2+W_y^2$ is proportional to the identity, so it can be absorbed into the
	energy parameter before the monodromy is formed. In that convention one obtains the block connection
	\begin{equation}
	\mathbb{A}_{\rm RD}(E;\varphi)=
	-\frac{\Phi}{\Phi_0}\mathbb{I}_4+
	\begin{pmatrix}
	W_\varphi(\varphi) & \mathbb{I}_2\\
	k^2(E)\mathbb{I}_2 & W_\varphi(\varphi)
	\end{pmatrix},
	\label{eq:RD_phase_space_connection}
	\end{equation}
	up to the harmless convention-dependent factors of $a$ already included in the definition
	\eqref{eq:Wphi_def}. Here $k^2(E)$ denotes the dimensionless energy after the scalar
	$W^2$ shift has been included. The important structural point is independent of this convention:
	the AB flux is central, the kinetic block carries the energy dependence, and the diagonal spin blocks
	carry the non-Abelian RD connection.
	The one-turn monodromy operator is therefore
	\begin{equation}
	\mathbb{U}_{2\pi}(E)=
	\mathcal{P}\exp\!\left(\int_0^{2\pi} d\varphi\;\mathbb{K}(E;\varphi)\right),
	\label{eq:RD_monodromy}
	\end{equation}
	or equivalently by using \eqref{eq:RD_covariant_transport} in the path-ordered exponential. The spectral
	problem is controlled by the boundary condition on the doubled state, not by $U_{\rm int}$ alone:
	\begin{equation}
	\det\!\left[\mathbb{U}_{2\pi}(E)-e^{-i\theta_0}\mathbb{I}_4\right]=0.
	\label{eq:RD_monodromy_quantization}
	\end{equation}
	In this sense the Wilson loop \eqref{eq:factor_RD} and the monodromy
	\eqref{eq:RD_monodromy_quantization} play distinct roles: the former organizes interferometric spin
	phases, while the latter contains the explicit energy dependence needed for quantization. A
	model-independent derivation of this phase-space doubling is recorded in Appendix~\ref{app:transport_holonomy}.
	
	\subsection{Curvature control of non-Abelian ordering}
	For generic $\alpha\neq\pm\beta$, the noncommutativity of $W_\varphi(\varphi)$ at different angles makes
	path ordering in \eqref{eq:factor_RD} essential. A controlled way to exhibit the algebraic content is
	the Magnus expansion,
	\begin{equation}
	U_{\rm int}=\exp\!\big(\Omega_1+\Omega_2+\cdots\big),
	\label{eq:Magnus}
	\end{equation}
	with
	\begin{align}
	\Omega_1 &= -i\int_0^{2\pi}\!d\varphi\;W_\varphi(\varphi),\\
	\Omega_2 &= -\frac{1}{2}\int_0^{2\pi}\!d\varphi_1\int_0^{\varphi_1}\!d\varphi_2\;
	\big[W_\varphi(\varphi_1),W_\varphi(\varphi_2)\big],
	\label{eq:Magnus_terms}
	\end{align}
	etc. The commutators appearing in $\Omega_2$ are the loop-space imprint of the non-Abelian curvature:
	for uniform couplings they are ultimately controlled by $F_{xy}$ in \eqref{eq:Fxy_RD}. In particular,
	as $\alpha\to\pm\beta$ the commutator contributions vanish, consistent with the pure-gauge interpretation.
	
	\subsection{Pure-gauge checkpoint and spin-helix symmetry}
	On the locus $\alpha=\pm\beta$ one has $F_{xy}=0$ by \eqref{eq:Fxy_RD}. In this case the $SU(2)$ connection
	is removable by a single-valued local spin rotation (for simply connected domains), and for the ring the
	resulting $SU(2)$ holonomy is conjugate to the identity,
	\begin{equation}
	U_{\rm int} \sim \mathbb{I},
	\qquad (\alpha=\pm\beta),
	\label{eq:U_puregauge}
	\end{equation}
	so that the loop physics reduces to the AB factor alone in \eqref{eq:factor_RD}. This is precisely the
	regime of exact $SU(2)$ symmetry and persistent-spin-helix behavior; it provides the natural ``null test''
	of the second composition: when the internal connection is pure gauge, the loop observable carries no
	intrinsic non-Abelian content.
	
	\subsection{Monodromy eigenphases and observable RD phases}
	For the RD ring, the spectral object is the energy-dependent monodromy
	\eqref{eq:RD_monodromy_quantization}. Its eigenvalue condition selects the allowed energies. The
	energy-independent internal holonomy \eqref{eq:factor_RD} is the corresponding geometric projection:
	it records the spin rotation accumulated along the loop and controls interference and spin filtering.
	Since $U_{\rm int}\in SU(2)$, its eigenvalues can be written as
	\begin{equation}
	\operatorname{spec}(U_{\rm int})=\left\{e^{+i\vartheta},e^{-i\vartheta}\right\},
	\qquad
	\vartheta\in[0,\pi],
	\label{eq:RD_eigenphases}
	\end{equation}
	with $\vartheta$ fixed by the conjugacy class of $U_{\rm int}$.
	The eigenphases of the full holonomy are therefore
	\begin{equation}
	\Theta_\pm(\Phi)=2\pi\frac{\Phi}{\Phi_0}\pm\vartheta
	\qquad (\mathrm{mod}\;2\pi),
	\label{eq:RD_total_phases}
	\end{equation}
	so the AB contribution appears as a central shift of the non-Abelian phase.
	This is the appropriate loop-space statement for interferometric observables. The corresponding spectral
	statement is obtained by applying the same phase matching to the eigenvalues of the doubled monodromy:
	\begin{equation}
	\lambda_j(E;\Phi)=e^{-i\theta_0},
	\qquad
	j=1,\ldots,4,
	\label{eq:RD_monodromy_eigenvalue_condition}
	\end{equation}
	which is equivalent to \eqref{eq:RD_monodromy_quantization}. In symmetry limits where the doubled
	monodromy factorizes into an orbital phase and an internal $SU(2)$ phase, this reduces to the transparent
	phase-matching rule
	\begin{equation}
	2\pi k(E)+2\pi\frac{\Phi}{\Phi_0}\pm\vartheta+\theta_0=2\pi n,
	\qquad n\in\mathbb{Z},
	\label{eq:RD_phase_matching}
	\end{equation}
	with the Berry/twist angle $\theta_0$ kept explicit rather than absorbed prematurely into an integer mode label.
	In the present manuscript \eqref{eq:RD_total_phases} is therefore used as the interferometric projection
	of the full transport problem, while \eqref{eq:RD_monodromy_quantization} is the spectral quantization
	condition. Closed-form expressions for $\vartheta$ in the pure-Rashba ($\beta=0$) and pure-Dresselhaus ($\alpha=0$)
	limits are given in Appendix~\ref{app:RD_limits}, while for general $\alpha,\beta$ the Magnus hierarchy of
	Appendix~\ref{app:Magnus} controls the corrections to $\vartheta$ order by order in curvature.
	
	\subsection{Summary: RD transport, holonomy, and interferometry}
	Equations \eqref{eq:min_coupling_SU2}--\eqref{eq:RD_monodromy_quantization} show that the RD problem on a
	ring has two nested geometric objects. The internal Wilson loop organizes spin rotation, interference, and
	spin filtering through the conjugacy class of $U_{\rm tot}(\Phi)$ and traced Wilson loops
	$\mathrm{Tr}\,U_{\rm tot}$. The doubled monodromy is the stronger object needed for spectra because it
	retains the energy dependence of the original Schr\"odinger problem. The non-Abelian character of both
	objects is measured by the curvature \eqref{eq:Fxy_RD} and the associated ordering corrections
	\eqref{eq:Magnus_terms}. This makes the RD ring the canonical condensed-matter application for the layered
	upgrade developed here: effective connection, energy-dependent transport, monodromy/holonomy, and finally
	observable flux and spin-interference data.
	
	\section{Geometric interpretation and diagrammatic construction}
	\label{sec:geometry_diagrams}
	
	\subsection{Two compositions as a single functorial pipeline}
	The formal development of Sec.~\ref{sec:loop_algebra}--Sec.~\ref{sec:RD_ring} can be summarized as a
	\emph{two-stage composition} that upgrades the original quantum-matter problem into loop-space data:
	
	\begin{equation}
	\begin{gathered}
	\text{quantum matter Hamiltonian}\\[-1pt]
	\xrightarrow{\;\mathfrak{C}_1\;}\\[-1pt]
	\substack{\text{effective }\Uone\text{ plus internal}\\
	\text{non-Abelian connection }\cA}\\[-1pt]
	\xrightarrow{\;\mathfrak{H}_{\cA}\;}\\[-1pt]
	\text{holonomies / loop observables}
	\end{gathered}
	\label{eq:two_compositions}
	\end{equation}
	
	The first map $\mathfrak{C}_1$ is the SOC-to-connection encoding (Sec.~\ref{sec:dirac_ring},
	Sec.~\ref{sec:RD_ring}); the second map $\mathfrak{H}_\cA$ is the holonomy assignment
	(transport functor) defined in Sec.~\ref{sec:loop_algebra}. The key structural point is that the
	second map preserves the composition law of paths:
	\begin{equation}
	\begin{aligned}
	&U(\gamma_2\circ\gamma_1)=U(\gamma_2)\,U(\gamma_1),
	\quad
	U(\gamma^{-1})=U(\gamma)^{-1},\\
	&U(\gamma)^\dagger=U(\gamma^{-1}),
	\end{aligned}
	\end{equation}
	so that the \emph{path/loop composition algebra} is represented as a \emph{$\ast$-algebra of unitary
	operators} on the internal Hilbert space. This is the precise sense in which the ``second composition''
	preserves the quantum algebraic structure.
	
\begin{figure}[t]
	\centering
	\includegraphics[width=\textwidth]{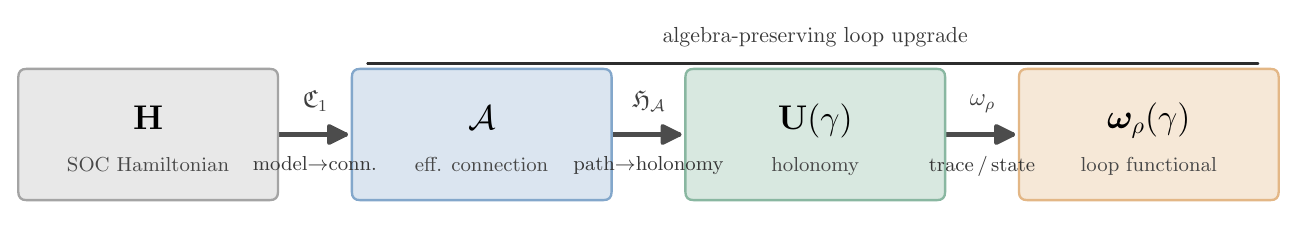}
	\caption{(Color online) Two-stage upgrade from the SOC Hamiltonian to effective connection data, then to
	holonomy and loop functionals. The second step preserves path composition and the $\ast$-structure.}
	\label{fig:composition}
	\end{figure}

	Figure~\ref{fig:composition} displays the two-stage map \eqref{eq:two_compositions} as a pipeline diagram
	$H \mapsto \cA \mapsto U(\gamma) \mapsto \omega_\rho(\gamma)=\Tr(\rho\,U(\gamma))$,
	where $\omega_\rho$ is a positive functional (state) on the holonomy algebra, as in
	\eqref{eq:state_functional}. The loop-representation viewpoint---reformulating gauge theories in terms
	of holonomies/Wilson loops---is classical in the loop calculus literature. \cite{Gambini1994,GambiniPullinBook}
	
	\subsection{Geometry on \texorpdfstring{$S^1$}{S1}: AB factor and internal holonomy}
	For ring geometries $M=S^1$, the fundamental group is $\pi_1(S^1)\cong\mathbb{Z}$, so the loop content is
	generated by a single class of windings. In this case the total $\Uone$ plus internal non-Abelian holonomy factorizes
	whenever the $\Uone$ part is central (proportional to the identity), as in Sec.~\ref{sec:dirac_ring} and
	Sec.~\ref{sec:RD_ring}:
	\begin{equation}
	\begin{aligned}
	U_{\rm tot}(\Phi)
	&=
	\exp\!\left(i2\pi\frac{\Phi}{\Phi_0}\right)\,
	U_{\rm int},\\
	U_{\rm int}
	&=\mathcal{P}\exp\!\left(-i\oint_{S^1}\cA_{\rm int}\right).
	\end{aligned}
	\label{eq:ring_factorization_geom}
	\end{equation}
	Thus AB flux enters as a commuting $\Uone$ phase, while SOC enters through a generally non-Abelian internal
	holonomy $U_{\rm int}$. Interferometric observables, and in spectral problems the corresponding monodromy data,
	are controlled by the conjugacy class of $U_{\rm tot}$ (equivalently its eigenvalues, or traced Wilson loops).
	
\begin{figure}[!t]
	\centering
	\includegraphics[width=0.72\textwidth]{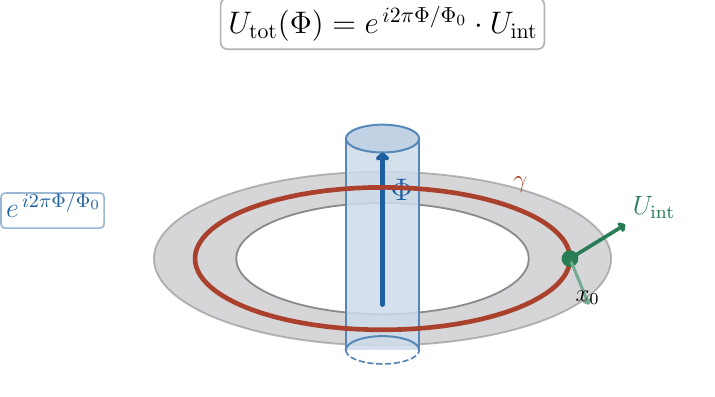}
	\caption{(Color online) Ring holonomy factorization.
	The AB sector contributes a central phase, while SOC induces internal transport based at $x_0$.
	When the $\Uone$ factor is central, the total loop datum factorizes as
	$U_{\rm tot}(\Phi)=e^{i2\pi\Phi/\Phi_0}\,U_{\rm int}$.}
	\label{fig:ring_holonomy}
	\end{figure}

	Figure~\ref{fig:ring_holonomy} illustrates this factorization: the central flux tube provides the $U(1)$ phase
	while the ``internal'' fiber rotation encodes the non-Abelian $U_{\rm int}$, with both contributions identified
	explicitly in \eqref{eq:ring_factorization_geom}.
	
	\subsection{Loop observables and reconstruction logic}
	Loop-space language is not only a visualization device: it provides a \emph{complete} set of gauge-covariant
	data for connections, at least under mild regularity assumptions. In particular, Wilson loops can be used to
	reconstruct gauge potentials up to gauge transformations in broad settings (reconstruction theorems). \cite{Giles1981}
	This motivates the ``second composition'' as a mathematically faithful reformulation: rather than treating
	$\cA$ as the primary object, one may take the holonomies $\{U(\gamma)\}$ (or traced loops $\{W(\gamma)\}$)
	as the fundamental geometric data.
	
	For the condensed-matter setting, we emphasize that $\cA$ is an effective/background connection fixed by
	material parameters and external fields; nonetheless, the map $\gamma\mapsto U(\gamma)$ retains its full
	algebraic content and provides a natural organizing principle for interference, and for spectral problems when a
	transport reduction is available, on multiply
	connected geometries.
	
	\subsection{Surface lift: curvature on spanning surfaces and the non-Abelian Stokes theorem}
	To connect loop pictures with surface-based diagrams (and to visualize non-Abelianity as ``curvature flux''),
	it is useful to lift the loop holonomy to a surface expression.
	Let $\gamma=\partial\Sigma$ bound an oriented surface $\Sigma$ and define the curvature
	\begin{equation}
	\cF=d\cA - i\,\cA\wedge\cA.
	\end{equation}
	A non-Abelian Stokes theorem rewrites the Wilson loop in terms of a surface-ordered exponential of
	(curvature dressed by parallel transport). Multiple equivalent formulations exist; see, e.g.,
	the operator/path-integral reviews. \cite{Broda2000,Kondo2008}
	
	A schematic form (suppressing surface ordering details) is
	\begin{equation}
	U(\gamma)
	=
	\mathcal{P}_\Sigma\exp\!\left(
	-i\int_{\Sigma} \widetilde{\cF}
	\right),
	\label{eq:NAST_schematic}
	\end{equation}
	where $\widetilde{\cF}$ is the curvature transported to a common reference point on $\Sigma$.
	Equation \eqref{eq:NAST_schematic} makes the geometric slogan precise:
	\emph{holonomy lives on loops, curvature lives on surfaces}.  In Sec.~\ref{sec:RD_ring} the
	non-Abelianity of Rashba--Dresselhaus transport is controlled by commutators, and therefore by curvature;
	the surface lift provides a direct way to visualize why the locus $\alpha=\pm\beta$ (vanishing curvature)
	is a ``pure gauge'' checkpoint.
	
\begin{figure}[tb]
	\centering
	\includegraphics[width=\textwidth]{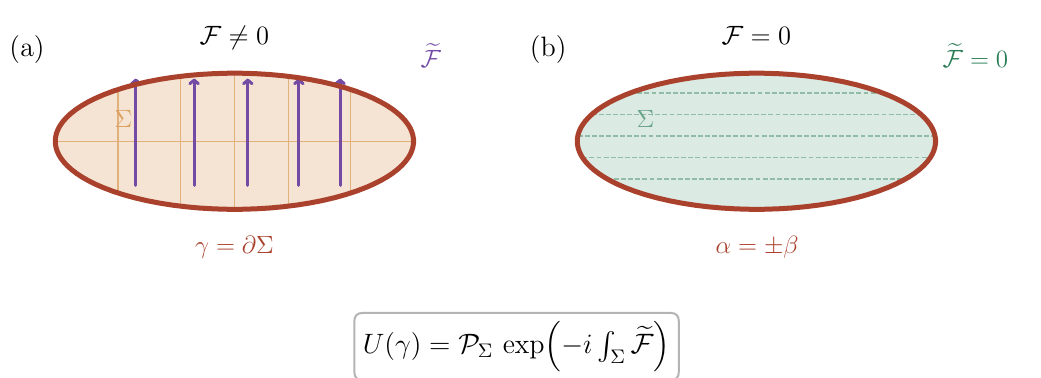}
	\caption{(Color online) Surface lift of the Wilson loop.
	The loop holonomy is rewritten as a surface-ordered exponential of transported curvature.
	The RD locus $\alpha=\pm\beta$ is a pure-gauge checkpoint with vanishing curvature. \cite{Broda2000,Kondo2008}}
	\label{fig:surface_lift}
	\end{figure}

	Figure~\ref{fig:surface_lift} depicts the loop $\gamma$ with spanning surface $\Sigma$: in the Abelian case
	curvature piercing $\Sigma$ reduces to the familiar flux picture; in the non-Abelian case the new element
	is surface ordering, encoded by the commutator structure of the Magnus expansion (Sec.~\ref{sec:RD_ring}).
	
	\subsection{Topological sectors and multivaluedness on multiply connected spaces}
	On $M=S^1$ the classification of sectors is governed by $\pi_1(S^1)\cong\mathbb{Z}$.
	In holonomy language, different windings probe powers of $U_{\rm tot}$:
	\begin{equation}
	U_{\rm tot}^{(n)}(\Phi)=\big[U_{\rm tot}(\Phi)\big]^n,
	\qquad n\in\mathbb{Z}.
	\label{eq:winding_holonomy}
	\end{equation}
	This is the loop-space expression of topological ``multivaluedness'' (monodromy): the state may acquire a
	nontrivial group element upon winding around the noncontractible cycle. In loop-dependent formulations,
	this is naturally encoded as a dependence of amplitudes on the loop class and its winding number. \cite{Gambini1994,GambiniPullinBook}
	In the present condensed-matter applications, the AB contribution gives the familiar $\Uone$ monodromy,
	while SOC contributes an internal non-Abelian monodromy; their product is the fundamental geometric datum
	organizing ring spectra and interference.
	
	\subsection{Geometric figures}
	The following figures illustrate the key stages of the holonomy construction.
	
	\subsection{Abelian and non-Abelian holonomy: complementary figures}
\begin{figure}[tb]
	\centering
	\includegraphics[width=0.95\textwidth]{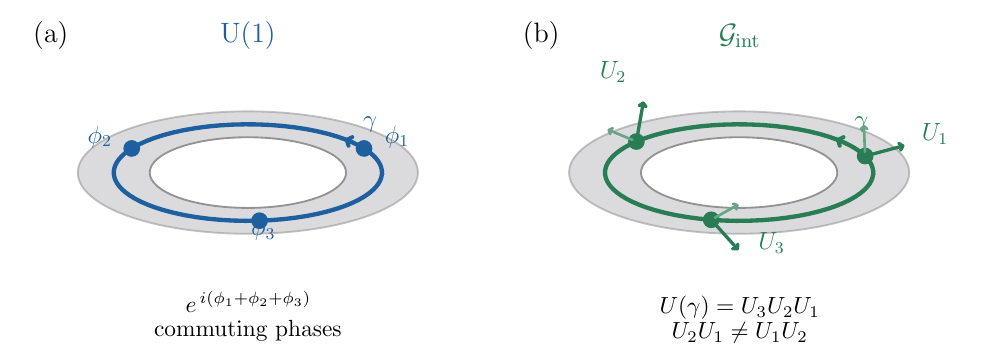}
	\caption{(Color online) Abelian versus non-Abelian transport around the same loop.
	The $\Uone$ sector reduces to a scalar phase, whereas the internal non-Abelian sector is an ordered product
	of segment propagators.}
	\label{fig:abelian_nonabelian}
	\end{figure}

\begin{figure}[tb]
	\centering
	\includegraphics[width=0.62\textwidth]{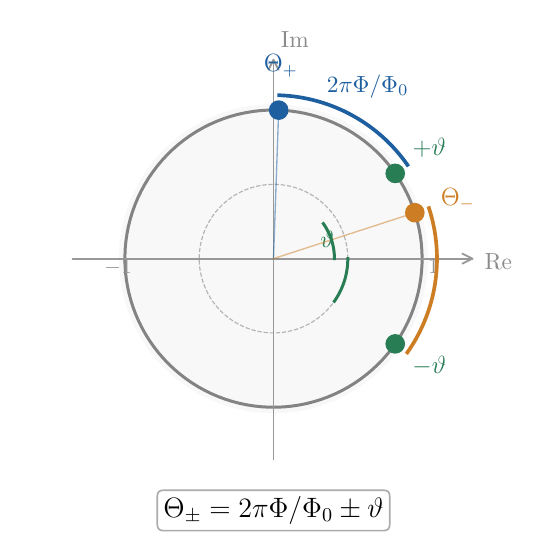}
	\caption{(Color online) Eigenphase representation of the conjugacy class.
	Internal phases $\pm\vartheta$ are shifted by the central AB contribution, yielding total phases $\Theta_\pm$.}
	\label{fig:eigenphase_circle}
	\end{figure}

\begin{figure}[tb]
	\centering
	\includegraphics[width=\textwidth]{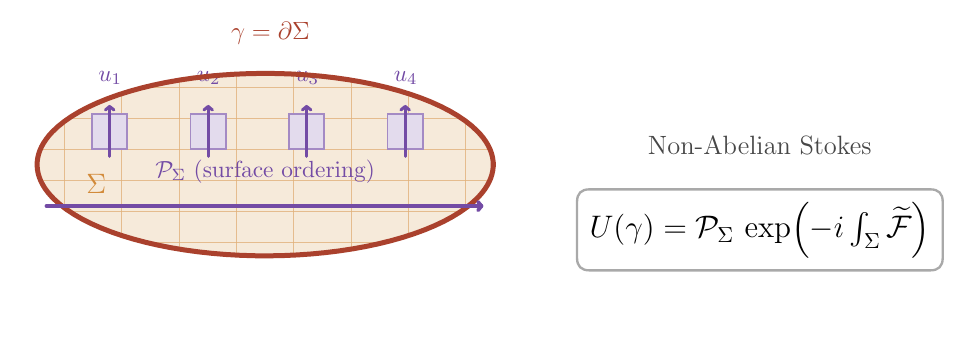}
	\caption{(Color online) Surface sweep and ordering in the non-Abelian Stokes picture.
	The spanning surface is decomposed into ordered strips transported to a common reference point.
	In Abelian problems the ordering is inessential; in the non-Abelian case it carries the commutator structure.}
	\label{fig:surface_sweep}
	\end{figure}

	Figures~\ref{fig:abelian_nonabelian}, \ref{fig:eigenphase_circle}, and \ref{fig:surface_sweep}
	isolate the ingredients that become nontrivial in the non-Abelian setting:
	ordered segment transport, conjugacy-class/eigenphase data, and surface ordering.
	
	\paragraph{Construction rule (practical).}
	In all applications, start from the effective connection in a chosen gauge/frame, compute the generator
	holonomy around the noncontractible loop(s), and reduce physical predictions to gauge-invariant data:
	eigenvalues of $U_{\rm tot}$, traced Wilson loops, or (for surfaces) curvature-induced surface ordering.
	This yields a uniform graphical language for AB phases, SOC-induced non-Abelian phases, and the special
	pure-gauge checkpoints.
	
\begin{figure}[ht]
\centering
\includegraphics[width=0.88\textwidth]{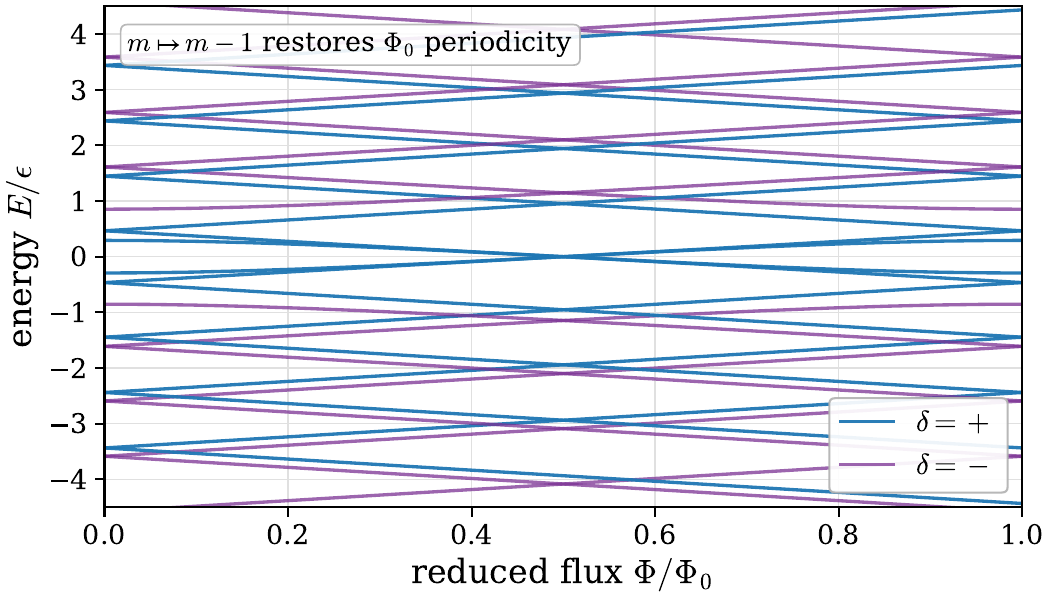}
\caption{(Color online) Monodromy spectrum of the Rashba graphene ring as a function of reduced flux
$\Phi/\Phi_0$. The AB sector enters as a central winding shift
$\tilde m=m+\Phi/\Phi_0$, while the Rashba coupling splits the internal spin/pseudospin branches.
The plot illustrates that spectral information is carried by the energy-dependent monodromy condition,
not by an energy-independent Wilson loop alone. The displayed parameters are
$\epsilon=1$, $\lambda_R=0.28$, and $m=-4,\ldots,4$.}
\label{fig:monodromy_spectrum_diagnostic}
\end{figure}

\begin{figure}[ht]
\centering
\includegraphics[width=0.74\textwidth]{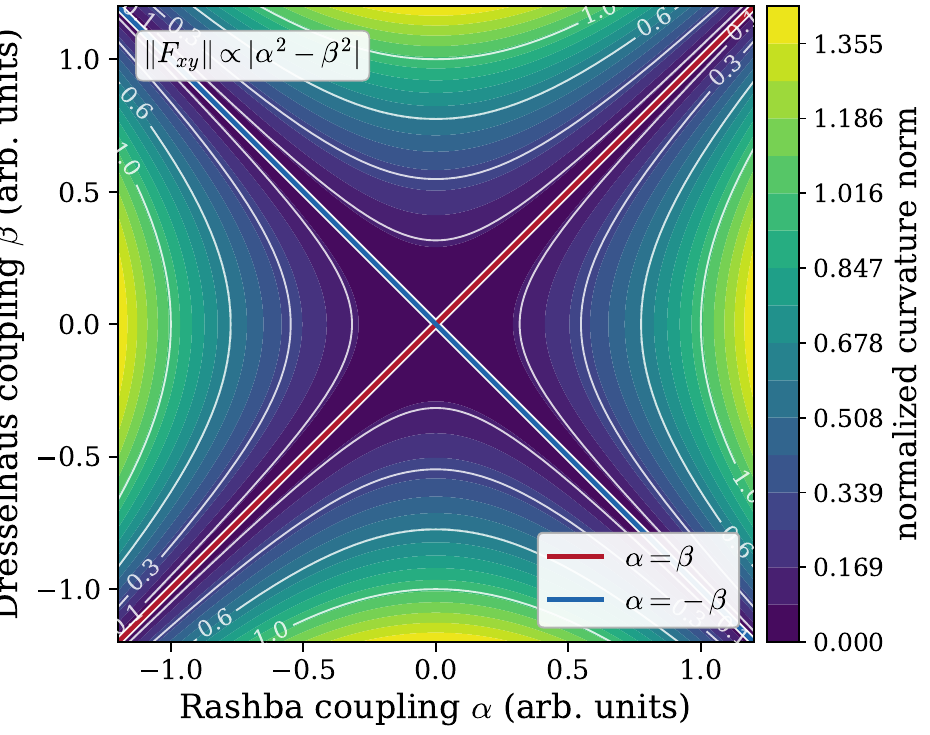}
\caption{(Color online) Non-Abelian curvature diagnostic for the Rashba--Dresselhaus connection.
The normalized curvature norm is proportional to $|\alpha^2-\beta^2|$ and vanishes on the pure-gauge
lines $\alpha=\pm\beta$. These lines provide a null test for the holonomy construction: local
commutator curvature disappears, while possible global ring twists must still be treated through
boundary monodromy.}
\label{fig:rd_curvature_diagnostic}
\end{figure}

	\paragraph{Validation-generated geometric diagnostics.}
	Two additional plots make the same distinction in a quantitative way.
	Figure~\ref{fig:monodromy_spectrum_diagnostic} displays the graphene-ring spectrum obtained from the
	energy-dependent monodromy condition, while Fig.~\ref{fig:rd_curvature_diagnostic} displays the
	Rashba--Dresselhaus curvature norm and its pure-gauge null lines.
	Both figures are generated by \texttt{validation/generate\_geometric\_diagnostics.py}; they are intended
	as geometric diagnostics, not as conductance simulations.
	
	\newpage
	
	\section{Conclusions and outlook}
	\label{sec:conclusion}
	
	\subsection{Conclusions}
	We have introduced a structure-preserving ``second composition'' that reformulates spin--orbit quantum matter in terms of loop-space
	data. The construction proceeds by (i) encoding the spin--orbit Hamiltonian as an effective $\Uone$ connection plus an internal non-Abelian connection, and
	(ii) mapping paths to holonomies $U(\gamma)$, thereby transferring the path/loop composition law into a $\ast$-algebra of unitary
	operators acting on the internal Hilbert space. In this sense, the loop upgrade preserves the quantum algebraic structure while
	replacing local gauge potentials by gauge-covariant holonomy data.
	
	In the Dirac (graphene) ring, the loop viewpoint yields a transparent unification of electromagnetic and internal transport:
	AB flux contributes a central $U(1)$ factor, while Rashba coupling contributes an internal holonomy. Because the Dirac ring is already
	first order, the same ordered transport object carries the energy dependence needed for a monodromy condition; this recasts the spectrum
	as a holonomy-eigenvalue problem and isolates the flux dependence as a universal winding-phase shift. This provides a compact geometric
	explanation for how interference and spectral flow are organized by a single energy-dependent transport datum.
	
	In the Rashba--Dresselhaus ring, the non-Abelian character is unavoidable: the internal connection generically rotates in spin space and
	fails to commute at different points on the loop. We showed how curvature controls this noncommutativity and singled out the pure-gauge
	locus $\alpha=\pm\beta$ as a stringent checkpoint where ordering becomes trivial up to conjugation. The Magnus expansion supplies a
	systematic commutator hierarchy for ordering effects. The non-Abelian Stokes theorem is used more modestly: it gives a surface-language
	representation of the same loop holonomy, useful for visualization and for identifying curvature/commutator content, but it does not
	replace the one-dimensional transport reduction that determines spectra.
	
	\subsection{Outlook: beyond a single ring}
	The construction developed here is deliberately formulated in a way that is
	not tied to a single circular geometry.
	
	\paragraph{(i) Multi-loop networks and graph topology.}
	A natural next step is to replace $S^1$ by a graph $\Gamma$ with several
	independent cycles. The fundamental group then changes from
	$\pi_1(S^1)\cong\mathbb{Z}$ to a free group generated by multiple loops, and
	the corresponding holonomy data become a noncommuting set of loop variables.
	This is the setting in which the geometric representation becomes more useful
	than a single-mode diagonalization: interference is organized by the
	composition algebra of paths rather than by one winding number.
	
	\paragraph{(ii) Other reconstructible spin systems.}
	A second extension is to apply the same reconstruction logic to spin
	Hamiltonians admitting effective gauge descriptions, including spin textures,
	synthetic spin--orbit couplings, and engineered non-Abelian transport in
	quantum devices. The framework does not require the effective connection to be
	a fundamental gauge field; it only requires that the physical Hamiltonian
	define a consistent transport problem.
	
	\paragraph{(iii) Holonomic control as a possible application.}
	Finally, the internal holonomy $U_{\rm int}$ may be interpreted as a
	spin-rotation gate generated by geometric transport. This suggests possible
	links with holonomic control protocols, although no quantum-information
	implementation is claimed here. The present work provides the transport
	geometry; questions of adiabaticity, noise, control fidelity, and device
	realization belong to a separate analysis.
	
	\paragraph{Final message.}
	The central claim is that spin--orbit Hamiltonians admitting an effective gauge
	reconstruction can be reformulated in terms of loop-observable data. In this
	formulation, holonomy organizes geometric and interferometric information,
	whereas energy-dependent monodromy carries spectral quantization. The second
	composition makes this principle algebraically precise without importing
	literal spin-network dynamics.
	
	\paragraph{Verification of algebraic identities.}
	The key algebraic identities---curvature signs, central factorization, comoving-frame conventions,
	and doubled transport structure---have been verified by direct computation using symbolic and
	numerical checks on finite-dimensional matrix representations.
	

\begin{appendix}

	\section{Transport form and holonomy eigenvalue condition}
	\label{app:transport_holonomy}
	
	\subsection{\texorpdfstring{General monodromy quantization on $S^1$}{General monodromy quantization on S1}}
	Let $M=S^1$ be parametrized by $\varphi\in[0,2\pi]$, and let the internal Hilbert space be
	$\mathcal{H}_{\rm int}$ (spin/pseudospin/valley, depending on the model).
	Suppose the stationary eigenvalue problem can be cast as a first-order transport equation
	\begin{equation}
	\partial_\varphi \chi(\varphi)= i\,\mathcal{A}_\varphi(E;\Phi,\ldots)\,\chi(\varphi),
	\label{eq:appA_transport_general}
	\end{equation}
	where $\mathcal{A}_\varphi$ is a Hermitian matrix acting on $\mathcal{H}_{\rm int}$ and may depend on
	the eigenvalue $E$ and on control parameters (flux $\Phi$, SOC strengths, etc.).
	
	Equation \eqref{eq:appA_transport_general} has the unique solution
	\begin{equation}
	\begin{aligned}
	\chi(\varphi)&=U(\varphi,0;E)\,\chi(0),\\
	U(\varphi,0;E)
	&=\mathcal{P}\exp\!\left(
	i\int_0^\varphi d\varphi'\,\mathcal{A}_\varphi(E)
	\right),
	\end{aligned}
	\label{eq:appA_Uphi}
	\end{equation}
	with path ordering in $\varphi$.
	
	Define the one-turn monodromy (holonomy) operator
	\begin{equation}
	U_{2\pi}(E)\equiv U(2\pi,0;E)=\mathcal{P}\exp\!\left(i\oint_{S^1}\mathcal{A}_\varphi(E)\,d\varphi\right).
	\label{eq:appA_holonomy}
	\end{equation}
	
	\subsection{Phase-space doubling for second-order ring equations}
	For Schr\"odinger-type ring problems, the stationary equation is naturally second order in the angular coordinate.
	A model-independent way to embed such problems into the transport form \eqref{eq:appA_transport_general}
	is to start from
	\begin{equation}
	\psi''(\varphi)+B_1(\varphi)\,\psi'(\varphi)+B_0(E;\varphi)\,\psi(\varphi)=0,
	\label{eq:appA_second_order_generic}
	\end{equation}
	where $\psi(\varphi)\in\mathcal{H}_{\rm int}$ and the matrix coefficients $B_1,B_0$ act on the internal space.
	Introducing the doubled state
	\begin{equation}
	\Psi(\varphi)\equiv
	\begin{pmatrix}
	\psi(\varphi)\\
	\psi'(\varphi)
	\end{pmatrix}
	\in
	\mathcal{H}_{\rm int}\oplus\mathcal{H}_{\rm int},
	\label{eq:appA_doubled_state}
	\end{equation}
	one obtains the equivalent first-order system
	\begin{equation}
	\begin{aligned}
	\partial_\varphi \Psi(\varphi)
	&=\mathbb{K}(E;\varphi)\,\Psi(\varphi),\\
	\mathbb{K}(E;\varphi)
	&=
	\begin{pmatrix}
	0 & \mathbb{I}\\
	-B_0(E;\varphi) & -B_1(\varphi)
	\end{pmatrix}.
	\end{aligned}
	\label{eq:appA_doubled_transport}
	\end{equation}
	The associated monodromy operator on the doubled space is
	\begin{equation}
	\mathbb{U}_{2\pi}(E)=
	\mathcal{P}\exp\!\left(\int_0^{2\pi}d\varphi\;\mathbb{K}(E;\varphi)\right).
	\label{eq:appA_doubled_monodromy}
	\end{equation}
	This is the correct object for spectral quantization in second-order ring problems: the purely internal
	$SU(2)$ Wilson loop extracted from the tangential connection captures geometric spin rotation, but the
	doubled monodromy \eqref{eq:appA_doubled_monodromy} is what carries the explicit energy dependence.
	
	\subsection{Boundary conditions and the eigenvalue condition}
	The physical boundary condition on a ring is, in general, a twisted condition
	\begin{equation}
	\chi(2\pi)=e^{-i\theta_0}\,\chi(0),
	\label{eq:appA_twistBC}
	\end{equation}
	where $\theta_0$ collects possible Berry/spin-connection contributions and/or convention-dependent frame
	rotations (for Dirac rings, this is where the half-integer shift information can sit).
	
	Combining \eqref{eq:appA_Uphi}--\eqref{eq:appA_twistBC} gives the monodromy condition
	\begin{equation}
	U_{2\pi}(E)\,\chi(0)=e^{-i\theta_0}\,\chi(0).
	\label{eq:appA_monodromy_condition}
	\end{equation}
	Nontrivial solutions exist iff $e^{-i\theta_0}$ is an eigenvalue of $U_{2\pi}(E)$, i.e.
	\begin{equation}
	\det\!\big(U_{2\pi}(E)-e^{-i\theta_0}\mathbb{I}\big)=0,
	\label{eq:appA_quant_condition}
	\end{equation}
	which is the holonomy-eigenvalue quantization condition used in the main text.
	
	\subsection{\texorpdfstring{Central $U(1)$ factor and AB shift}{Central U(1) factor and AB shift}}
	If the connection splits as
	\begin{equation}
	\mathcal{A}_\varphi(E;\Phi)=\left(\frac{\Phi}{\Phi_0}\right)\mathbb{I}+\mathcal{A}^{\rm int}_\varphi(E),
	\label{eq:appA_split}
	\end{equation}
	with the AB term proportional to $\mathbb{I}$, then it commutes with the internal sector and the holonomy
	factorizes exactly:
	\begin{equation}
	\begin{aligned}
	&U_{2\pi}(E;\Phi)=e^{\,i2\pi\Phi/\Phi_0}\;U_{2\pi}^{\rm int}(E),\\
	&U_{2\pi}^{\rm int}(E)=\mathcal{P}\exp\!\left(i\oint\mathcal{A}^{\rm int}_\varphi(E)\,d\varphi\right).
	\end{aligned}
	\label{eq:appA_factor}
	\end{equation}
	Thus AB flux enters the quantization condition \eqref{eq:appA_quant_condition} as a commuting phase factor,
	equivalently as the standard shift $m\mapsto m+\Phi/\Phi_0$ when one uses Fourier modes.
	
	\subsection{Graphene-ring specialization (link to Sec.~\ref{sec:dirac_ring})}
	In the graphene ring (Sec.~\ref{sec:dirac_ring}), after the comoving-frame rotation the eigenvalue problem
	reduces to \eqref{eq:appA_transport_general} with
	$\mathcal{A}_\varphi(E;\Phi)$ given explicitly by Eq.~\eqref{eq:Aphi_graphene}.
	Since $\mathcal{A}_\varphi$ is $\varphi$-independent there, path ordering is trivial and
	$U_{2\pi}(E;\Phi)=\exp\!\big(i2\pi\mathcal{A}_\varphi(E;\Phi)\big)$, recovering Eq.~\eqref{eq:W_graphene}
	and the factorization Eq.~\eqref{eq:factor_graphene}.
	
	\subsection{Nonrelativistic ring specialization (link to Sec.~\ref{sec:RD_ring})}
	In the Rashba--Dresselhaus ring (Sec.~\ref{sec:RD_ring}), the internal Wilson loop takes the form
	\eqref{eq:appA_holonomy}, but the spectral problem is governed by the phase-space doubled transport
	\eqref{eq:appA_second_order_generic}--\eqref{eq:appA_doubled_monodromy}. The relevant monodromy therefore
	acts on $\mathcal{H}_{\rm int}\oplus\mathcal{H}_{\rm int}$ and carries the energy dependence inherited from
	the original Schr\"odinger equation. The AB contribution remains central and factorizes as in
	Eq.~\eqref{eq:factor_RD}. Thus the $SU(2)$ factor $U_{\SU(2)}$ and the eigenphases
	\eqref{eq:RD_total_phases} should be read as phase/interference data, while the spectral condition is the
	doubled determinant \eqref{eq:RD_monodromy_quantization}.
	
	\section{Non-Abelian Stokes theorem and surface-ordering conventions}
	\label{app:NAST}
	
	\subsection{Scope and why conventions matter}
	Unlike the Abelian Stokes theorem, there is no unique, universal ``non-Abelian Stokes theorem'' (NAST) formula
	because Lie-algebra valued quantities do not commute and one must specify (i) an ordering prescription and
	(ii) a rule for comparing algebra elements at different points on a surface.
	A useful, standard strategy is to express a Wilson loop as a \emph{surface-ordered} exponential of a
	\emph{parallel-transported curvature} (operator/product-integral form), or alternatively as a
	\emph{path-integral/coherent-state} representation (Diakonov--Petrov type), which replaces ordering by an
	auxiliary integration. See Broda's review for a clear taxonomy of these approaches. \cite{Broda2000} 
	We adopt the operator/product-integral conventions below, since they are the most directly compatible with the
	geometric/diagrammatic narrative in Sec.~\ref{sec:geometry_diagrams} and with the curvature/commutator control
	of Sec.~\ref{sec:RD_ring}. \cite{Broda2000,KarpMansouriRno1999}
	
	\subsection{Wilson lines, Wilson loops, and path ordering}
	Let $G$ be a compact Lie group with Lie algebra $\mathfrak{g}$ and let $\cA=\cA_\mu(x)\,dx^\mu$ be a
	$\mathfrak{g}$-valued connection one-form in a fixed unitary representation on $\mathcal{H}_{\rm int}$.
	For an oriented path $\gamma:[0,1]\to M$, the Wilson line (parallel transporter) is
	\begin{equation}
	U[\gamma]\equiv U_\gamma[\cA]
	:= \cP \exp\!\left(-i\int_0^1 ds\;\dot{\gamma}^\mu(s)\,\cA_\mu(\gamma(s))\right),
	\label{eq:U_path_def_app}
	\end{equation}
	where $\cP$ denotes path ordering (increasing $s$ to the left).
	For a closed loop $C$ based at $x_0$ (i.e.\ $C(0)=C(1)=x_0$), the Wilson loop operator in representation $R$ is
	\begin{equation}
	W_R[C] := \Tr_R\,U[C].
	\label{eq:Wloop_def_app}
	\end{equation}
	These definitions match the standard usage in NAST expositions and in the DP-type treatments. \cite{Broda2000,Kondo2008,HirayamaUeno2000}
	
	\subsection{Curvature and the need for ``dressing''}
	The curvature two-form is
	\begin{equation}
	\begin{aligned}
	&\cF = d\cA - i\,\cA\wedge \cA
	=\frac12\,\cF_{\mu\nu}(x)\,dx^\mu\wedge dx^\nu,\\
	&\cF_{\mu\nu}=\partial_\mu\cA_\nu-\partial_\nu\cA_\mu-i[\cA_\mu,\cA_\nu].
	\end{aligned}
	\label{eq:curvature_app}
	\end{equation}
	A naive attempt to write $U[C]\stackrel{?}{=}\exp(-i\int_\Sigma \cF)$ fails because $\cF(x)$ at distinct points
	does not commute. A second obstruction is that $\cF(x)$ lives in the fiber at $x$, so one must specify how
	to compare $\cF(x)$ for different $x\in\Sigma$.
	The standard cure is to ``dress'' the curvature by parallel transport to a common reference point.
	
	\subsection{Surface scanning and surface ordering}
	Let $\Sigma$ be an oriented surface with boundary $\partial\Sigma=C$ and let $x_0\in C$ be the base point.
	Choose a smooth parameterization $\Sigma(u,v)$ with $(u,v)\in[0,1]\times[0,1]$, such that:
	\begin{itemize}
	\item $\Sigma(0,v)=x_0$ for all $v$ (a convenient ``spine'' anchored at the base point),
	\item $\Sigma(1,v)$ traces the boundary $C$ (up to reparameterization),
	\item for fixed $u$, the curve $v\mapsto \Sigma(u,v)$ lies entirely in $\Sigma$ and defines a family of
	      partial loops $C_u$ that ``sweep'' the surface as $u$ increases from $0$ to $1$.
	\end{itemize}
	This ``surface scanning'' choice is part of the convention: different scans lead to equivalent results,
	but the intermediate expressions differ by reorganization of ordering. Broda discusses these issues in detail
	and emphasizes that multiple NAST variants exist. \cite{Broda2000}
	
	Define the tangent vectors
	\begin{equation}
	\partial_u x^\mu(u,v):=\frac{\partial \Sigma^\mu}{\partial u},\qquad
	\partial_v x^\mu(u,v):=\frac{\partial \Sigma^\mu}{\partial v}.
	\end{equation}
	The induced area element is $dS^{\mu\nu}=\left(\partial_u x^\mu \partial_v x^\nu-\partial_u x^\nu \partial_v x^\mu\right)\,du\,dv$.
	
	\subsection{Parallel-transported curvature}
	For each point $x=\Sigma(u,v)$ pick a canonical path within $\Sigma$ from $x_0$ to $x$, e.g.\ first move along the
	$u$-direction at $v=0$ and then along the $v$-direction at fixed $u$:
	\begin{equation}
	\ell_{(u,v)}: x_0=\Sigma(0,0)\to \Sigma(u,0)\to \Sigma(u,v)=x.
	\end{equation}
	Let $U[\ell_{(u,v)}]$ be the Wilson line along this path. The dressed curvature at $x$ is
	\begin{equation}
	\widetilde{\cF}_{\mu\nu}(u,v)
	:= U[\ell_{(u,v)}]^{-1}\,\cF_{\mu\nu}(\Sigma(u,v))\,U[\ell_{(u,v)}].
	\label{eq:dressed_curv_app}
	\end{equation}
	This conjugation makes $\widetilde{\cF}$ an algebra element ``based'' at $x_0$ so that contributions from
	different surface points can be consistently ordered and multiplied.
	This dressing is the precise operator-level version of the ``curvature transported to a common reference point''
	statement in Sec.~\ref{sec:geometry_diagrams}. \cite{Broda2000,KarpMansouriRno1999}
	
	\subsection{Operator/product-integral non-Abelian Stokes theorem}
	With the above conventions, one form of the operator NAST reads
	\begin{equation}
	\begin{aligned}
	U[C]
	&=\cP_\Sigma\exp\!\left(
	-i\int_0^1 du\int_0^1 dv\;\widetilde{\cF}_{\mu\nu}(u,v)\right.\\
	&\qquad\left.\times\partial_u x^\mu(u,v)\,\partial_v x^\nu(u,v)
	\right),
	\end{aligned}
	\label{eq:NAST_operator_app}
	\end{equation}
	where $\cP_\Sigma$ is a \emph{surface-ordering} operator induced by the scan parameter $u$:
	contributions at larger $u$ are ordered to the left (analogous to time ordering),
	while the $v$ integration at fixed $u$ is understood as producing an infinitesimal ordered contribution to the
	evolution in $u$. This operator form and its variants are reviewed and derived in the operator approach discussed
	by Broda, and in product-integral proofs such as Karp--Mansouri--Rno. \cite{Broda2000,KarpMansouriRno1999}
	
	\paragraph{Discrete (plaquette) definition of $\cP_\Sigma$.}
	A practical and unambiguous way to define $\cP_\Sigma$ is via discretization. Tile $\Sigma$ into small plaquettes
	$\{\Delta S_k\}$ ordered by increasing $u$, and approximate
	\begin{equation}
	U[C]
	\approx
	\prod_{k=N}^{1}\exp\!\big(-i\,\widetilde{\cF}(x_k)\,\Delta S_k\big),
	\label{eq:plaquette_order_app}
	\end{equation}
	where $x_k$ is a point in plaquette $k$ and the product is ordered with $k=N$ (largest $u$) leftmost.
	The product-integral approach formalizes the continuum limit of \eqref{eq:plaquette_order_app} and provides
	clean proofs of the NAST under mild assumptions. \cite{KarpMansouriRno1999}
	
	\subsection{Gauge covariance}
	Under $g:M\to G$, $\cA\mapsto \cA^g=g\cA g^{-1}+i\,dg\,g^{-1}$ and $\cF\mapsto \cF^g=g\cF g^{-1}$.
	One checks that $U[\ell_{(u,v)}]\mapsto g(x_0)U[\ell_{(u,v)}]g(x)^{-1}$, hence
	$\widetilde{\cF}\mapsto g(x_0)\widetilde{\cF}\,g(x_0)^{-1}$, i.e.\ dressed curvature transforms by a \emph{single}
	conjugation at the base point. Therefore the surface-ordered exponential \eqref{eq:NAST_operator_app} transforms as
	\begin{equation}
	U[C]\mapsto g(x_0)\,U[C]\,g(x_0)^{-1},
	\end{equation}
	and traced Wilson loops $W_R[C]=\Tr_R U[C]$ are gauge invariant.
	These covariance properties are emphasized in NAST derivations (including DP-type formulations) because they ensure
	that the final surface expression computes the same gauge-invariant observable as the original path-ordered definition.
	\cite{Kondo2008,HirayamaUeno2000}
	
	\subsection{Abelian limit and commuting-curvature regime}
	If $G=\Uone$ (or if $\cA$ is restricted to an Abelian subalgebra), all commutators vanish, $\widetilde{\cF}=\cF$,
	surface ordering becomes irrelevant, and \eqref{eq:NAST_operator_app} reduces to the standard Abelian Stokes theorem:
	\begin{equation}
	U[C]=\exp\!\left(-i\int_\Sigma \cF\right).
	\end{equation}
	More generally, in regimes where $\big[\widetilde{\cF}(x),\widetilde{\cF}(y)\big]$ is negligible over the surface
	(e.g.\ sufficiently small surfaces or effectively commuting sectors), the leading approximation is Abelian-like and
	ordering corrections are systematically organized by commutators (Magnus-type expansions), matching the logic used
	in Sec.~\ref{sec:RD_ring}. \cite{Broda2000}
	
	\subsection{Diakonov--Petrov (coherent-state) type NAST (alternative form)}
	For some applications it is advantageous to trade ordering for an auxiliary integration over group variables on $\Sigma$.
	The Diakonov--Petrov (DP) type NAST rewrites the Wilson loop as a surface expression without explicit surface ordering,
	at the price of introducing a functional integral over group-valued fields (or coherent-state variables).
	Kondo provides a pedagogical derivation and discussion of DP-type NAST and its relation to gauge-invariant
	representations of the Wilson loop. \cite{Kondo2008}
	Hirayama--Ueno also present a path-integral formula applicable to general compact semi-simple gauge groups. \cite{HirayamaUeno2000}
	In this manuscript we use the operator/product-integral conventions \eqref{eq:NAST_operator_app}--\eqref{eq:plaquette_order_app}
	as our default, but DP-type formulas can be used interchangeably as cross-checks in non-Abelian settings.
	
	\subsection{Practical rule for the figures and for ring-based applications}
	When drawing (or computing) the ``surface lift'' of a loop observable, the following steps implement our conventions:
	\begin{enumerate}
	\item Pick a base point $x_0\in C$ and a scan of $\Sigma$ (family of partial loops $C_u$).
	\item Define reference paths $\ell_{(u,v)}$ within $\Sigma$ and compute dressed curvature $\widetilde{\cF}$ by \eqref{eq:dressed_curv_app}.
	\item Evaluate the surface-ordered exponential \eqref{eq:NAST_operator_app} either analytically (special symmetry/pure-gauge loci)
	or numerically by the plaquette product \eqref{eq:plaquette_order_app}.
	\end{enumerate}
	In the Rashba--Dresselhaus ring problem, noncommutativity enters through the curvature/commutator sector and is
	geometrically represented as ``curvature piercing $\Sigma$''; at $\alpha=\pm\beta$ the curvature vanishes and
	the surface lift becomes trivial up to conjugation, giving a clean null test for the non-Abelian structure.
	\cite{Broda2000,Kondo2008}
	
	\section{Magnus expansion, ordering effects, and curvature control}
	\label{app:Magnus}
	
	\subsection{Magnus expansion for path-ordered exponentials}
	Consider a linear transport equation on a Lie group (or unitary operators) of the form
	\begin{equation}
	\frac{d}{dt}U(t)=A(t)\,U(t),\qquad U(0)=\mathbb{I},
	\label{eq:Magnus_ODE}
	\end{equation}
	with $A(t)$ an (in general noncommuting) operator-valued generator. The formal solution is the
	time-/path-ordered exponential
	\begin{equation}
	U(t)=\mathcal{T}\exp\!\left(\int_0^t A(\tau)\,d\tau\right),
	\label{eq:ordered_exp}
	\end{equation}
	where $\mathcal{T}$ orders larger $\tau$ to the left. The Magnus expansion states that $U(t)$ can be written as a
	single exponential
	\begin{equation}
	U(t)=\exp\!\big(\Omega(t)\big),\qquad \Omega(t)=\sum_{n\ge 1}\Omega_n(t),
	\label{eq:Magnus_form}
	\end{equation}
	with the first terms
	\begin{align}
	\Omega_1(t) &= \int_0^t A(t_1)\,dt_1, \label{eq:Omega1}\\
	\Omega_2(t) &= \frac12\int_0^t dt_1\int_0^{t_1}dt_2\;[A(t_1),A(t_2)], \label{eq:Omega2}\\
	\Omega_3(t) &= \frac{1}{6}\int_0^t dt_1
	\int_0^{t_1}dt_2\int_0^{t_2}dt_3\notag\\
	&\quad\times
	\Big(
	[A(t_1),[A(t_2),A(t_3)]]\notag\\
	&\qquad\qquad
	+[A(t_3),[A(t_2),A(t_1)]]
	\Big),
	\label{eq:Omega3}
	\end{align}
	and so on. Each $\Omega_n$ is a nested-commutator functional of $A$, hence ordering effects are systematically
	organized by commutators. This is the standard tool in Lie-group integrators and quantum evolution, and preserves
	group structure order-by-order. \cite{Blanes2009PhysRep,Iserles2000ActaNum} 
	
	\paragraph{Convergence (practical bound).}
	A sufficient condition (not sharp) for convergence is that $\int_0^t\|A(\tau)\|\,d\tau<\pi$ in a suitable operator norm.
	For our usage, we primarily employ the Magnus series as a controlled expansion when commutator contributions are
	small (e.g., small loops/surfaces, nearly commuting sectors, or pure-gauge checkpoints). \cite{Blanes2009PhysRep}
	
	\subsection{From surface ordering to a Magnus expansion in the scan parameter}
	In Appendix~\ref{app:NAST} we adopted an operator/product-integral form of the non-Abelian Stokes theorem (NAST),
	\begin{equation}
	U[C]
	=
	\mathcal{P}_\Sigma\exp\!\left(
	-i\int_0^1 du\int_0^1 dv\;
	\widetilde{\cF}_{\mu\nu}(u,v)\;
	\partial_u x^\mu\,\partial_v x^\nu
	\right),
	\label{eq:NAST_operator_again}
	\end{equation}
	where $\widetilde{\cF}$ is the curvature dressed to a common base point by parallel transport (Appendix~\ref{app:NAST}),
	and $\mathcal{P}_\Sigma$ is induced by a surface scan parameter $u$ (larger $u$ ordered to the left). \cite{Broda2000,KarpMansouriRno1999}
	
	A convenient way to expose the algebra behind $\mathcal{P}_\Sigma$ is to define an \emph{effective $u$-generator}
	\begin{equation}
	B(u)\equiv -i\int_0^1 dv\;\widetilde{\cF}_{\mu\nu}(u,v)\;
	\partial_u x^\mu(u,v)\,\partial_v x^\nu(u,v),
	\label{eq:Bu_def}
	\end{equation}
	so that \eqref{eq:NAST_operator_again} becomes a \emph{$u$-ordered} exponential,
	\begin{equation}
	U[C]=\mathcal{T}_u\exp\!\left(\int_0^1 B(u)\,du\right).
	\label{eq:U_uordered}
	\end{equation}
	Equation \eqref{eq:U_uordered} is now precisely in the Magnus form \eqref{eq:ordered_exp}, with ``time'' $t\mapsto u$
	and generator $A(t)\mapsto B(u)$. Therefore,
	\begin{equation}
	U[C]=\exp\!\Big(\Omega_\Sigma\Big),\qquad \Omega_\Sigma=\sum_{n\ge 1}\Omega_{\Sigma,n},
	\label{eq:Magnus_surface}
	\end{equation}
	where
	\begin{align}
	\Omega_{\Sigma,1} &= \int_0^1 du\;B(u)\notag\\
	&= -i\int_0^1 du\int_0^1 dv\;
	\widetilde{\cF}_{\mu\nu}(u,v)\;
	\partial_u x^\mu\,\partial_v x^\nu,
	\label{eq:OmegaSigma1}\\[3pt]
	\Omega_{\Sigma,2} &= \frac12\int_0^1 du_1\int_0^{u_1}du_2\;[B(u_1),B(u_2)].
	\label{eq:OmegaSigma2}
	\end{align}
	Thus the \emph{first} term is the dressed-curvature flux through $\Sigma$, and the \emph{second} term is the leading
	surface-ordering correction, built from commutators of curvature fluxes in different scan slices. This is the precise
	sense in which ``ordering is controlled by curvature noncommutativity'' in our surface-lift diagrams. \cite{Blanes2009PhysRep,Broda2000}
	
	\subsection{Small-surface / commuting-curvature regime}
	If $\widetilde{\cF}(u,v)$ effectively lies in an Abelian subalgebra on $\Sigma$ (or if $\Sigma$ is small enough that
	commutators are negligible), then $[B(u_1),B(u_2)]\approx 0$ and $\Omega_{\Sigma,n\ge2}$ are suppressed:
	\begin{equation}
	U[C]\approx \exp\!\left(-i\int_\Sigma \widetilde{\cF}\right),
	\label{eq:abelian_like}
	\end{equation}
	recovering the familiar Abelian flux picture as a controlled approximation.
	The dominant corrections are $\mathcal{O}([\widetilde{\cF},\widetilde{\cF}])$ and are explicitly given by
	\eqref{eq:OmegaSigma2}--\eqref{eq:Omega3} with $B(u)$ from \eqref{eq:Bu_def}. \cite{Blanes2009PhysRep}
	
	\subsection{Connection to the loop (1D) Magnus expansion used in Sec.~\ref{sec:RD_ring}}
	In Sec.~\ref{sec:RD_ring} we wrote the $SU(2)$ ring holonomy as the path-ordered exponential
	\begin{equation}
	U_{\SU(2)}=\mathcal{P}\exp\!\left(-i\int_0^{2\pi}\!d\varphi\;W_\varphi(\varphi)\right),
	\label{eq:U_RD_again}
	\end{equation}
	and invoked the Magnus expansion
	\begin{equation}
	\begin{aligned}
	&U_{\SU(2)}=\exp(\Omega_1+\Omega_2+\cdots),\\
	&\Omega_2=-\frac12\int d\varphi_1\int^{\varphi_1} d\varphi_2\;[W_\varphi(\varphi_1),W_\varphi(\varphi_2)].
	\end{aligned}
	\label{eq:Magnus_ring_again}
	\end{equation}
	This is the 1D analog of the surface Magnus construction above: the generator is $A(\varphi)=-i\,W_\varphi(\varphi)$.
	
	To see explicitly how curvature enters, recall that for uniform Rashba--Dresselhaus couplings the planar
	curvature is purely commutator-generated,
	\begin{equation}
	F_{xy}=-i[W_x,W_y]\propto(\alpha^2-\beta^2)\sigma_z,
	\label{eq:Fxy_again}
	\end{equation}
	and $W_\varphi(\varphi)=a\,\hat{\bm e}_\varphi\!\cdot\!\bm W$ is a rotating linear combination of $W_x,W_y$.
	Therefore, the commutators in $\Omega_2$ are ultimately controlled by $[W_x,W_y]$ and hence by $F_{xy}$.
	In particular, on the pure-gauge locus $\alpha=\pm\beta$ one has $F_{xy}=0$ and the commutator sector vanishes,
	so $\Omega_{n\ge2}=0$ and ordering becomes trivial up to conjugation, reproducing the checkpoint discussed in
	Sec.~\ref{sec:RD_ring}. \cite{Broda2000,Blanes2009PhysRep}
	
	\subsection{Practical takeaway for computations and figures}
	The two Magnus expansions in this manuscript play complementary roles:
	\begin{itemize}
	\item \emph{Loop Magnus} (Sec.~\ref{sec:RD_ring}): organizes non-Abelian corrections to transport along $S^1$ via
	nested commutators of the tangential connection $W_\varphi$.
	\item \emph{Surface Magnus} (this Appendix): organizes surface-ordering corrections in the NAST via nested commutators
	of dressed-curvature ``slices'' $B(u)$.
	\end{itemize}
	Both make the same structural statement explicit: \emph{non-Abelianity enters through commutators}, and in turn
	commutators are geometrically encoded by curvature.
	This provides the clean logic behind our diagram set in Sec.~\ref{sec:geometry_diagrams}: in Abelian or pure-gauge
	limits, loops reduce to flux pictures; away from those limits, ordering corrections are controlled (and visualized)
	by curvature piercing spanning surfaces. \cite{Broda2000,KarpMansouriRno1999,Blanes2009PhysRep}

	
	\section{Explicit diagonalization for the Dirac (graphene) Rashba ring}
	\label{app:diag_graphene}
	
	\subsection{\texorpdfstring{Mode decomposition and $4\times 4$ Hamiltonian}{Mode decomposition and 4 x 4 Hamiltonian}}
	We start from the comoving-frame, flux-threaded ring Hamiltonian of Sec.~\ref{sec:dirac_ring}
	(we keep the same notation there). For each integer angular channel $m\in\mathbb{Z}$ we write
	\begin{equation}
	\Psi'(\varphi)=e^{im\varphi}\,\psi_m,
	\qquad
	\tilde m := m+\frac{\Phi}{\Phi_0},
	\qquad
	\epsilon := \frac{\hbar v_F}{a}.
	\end{equation}
	In the Rashba-only case (intrinsic SOC set to zero for clarity in the diagonalization), the
	Hamiltonian in the comoving frame acts on the four-component internal spinor
	$\psi_m\in\mathbb{C}^2_{\rm ps}\otimes\mathbb{C}^2_{\rm s}$ (pseudospin $\otimes$ spin) as
	\begin{equation}
	H_m
	=
	\epsilon\,\tilde m\,\sigma_y\otimes\mathbb{I}
	-\frac{\epsilon}{2}\,\sigma_y\otimes s_z
	+\lambda_R\left(\sigma_x\otimes s_y - \sigma_y\otimes s_x\right).
	\label{eq:appC_Hm_def}
	\end{equation}
	In the canonical basis
	\begin{equation}
	\big\{|A\uparrow\rangle,\,|A\downarrow\rangle,\,|B\uparrow\rangle,\,|B\downarrow\rangle\big\},
	\end{equation}
	$H_m$ has the chiral block form
	\begin{equation}
	H_m=
	\begin{pmatrix}
	0 & B_m\\
	B_m^\dagger & 0
	\end{pmatrix},
	\qquad
	B_m
	=
	i\begin{pmatrix}
	\frac{\epsilon}{2}-\epsilon\tilde m & 0\\[3pt]
	2\lambda_R & -\left(\epsilon\tilde m+\frac{\epsilon}{2}\right)
	\end{pmatrix}.
	\label{eq:appC_block_form}
	\end{equation}
	(One may check $H_m^\dagger=H_m$ explicitly.)
	
	\subsection{\texorpdfstring{Spectrum from the singular values of $B_m$}{Spectrum from the singular values of Bm}}
	Because of the off-diagonal structure \eqref{eq:appC_block_form},
	\begin{equation}
	H_m^2=
	\begin{pmatrix}
	B_mB_m^\dagger & 0\\
	0 & B_m^\dagger B_m
	\end{pmatrix},
	\end{equation}
	so the squared energies are the eigenvalues of the $2\times 2$ positive matrix $B_mB_m^\dagger$
	(and appear twice because $B_mB_m^\dagger$ and $B_m^\dagger B_m$ have the same nonzero spectrum).
	
	Write
	\begin{equation}
	A:=\epsilon\tilde m,\qquad b:=\frac{\epsilon}{2}.
	\end{equation}
	Up to the trivial factor $i$ in \eqref{eq:appC_block_form}, the relevant real matrix is
	\begin{equation}
	M_m :=
	\begin{pmatrix}
	b-A & 0\\
	2\lambda_R & -(A+b)
	\end{pmatrix},
	\quad
	B_m=i\,M_m,
	\end{equation}
	where $B_mB_m^\dagger=M_mM_m^{\mathsf T}$.
	A short calculation gives
	\begin{equation}
	M_mM_m^{\mathsf T}
	=
	\begin{pmatrix}
	(b-A)^2 & 2\lambda_R(b-A)\\
	2\lambda_R(b-A) & (A+b)^2+4\lambda_R^2
	\end{pmatrix}.
	\label{eq:appC_MM}
	\end{equation}
	The two eigenvalues $\mu_\delta$ ($\delta=\pm$) of \eqref{eq:appC_MM} are
	\begin{equation}
	\mu_\delta
	=
	\left(A^2+b^2\right)+2\lambda_R^2
	\;-\;2\delta\sqrt{\left(A^2+\lambda_R^2\right)\left(b^2+\lambda_R^2\right)}.
	\label{eq:appC_mu_delta}
	\end{equation}
	Therefore the energy eigenvalues are
	\begin{equation}
	E_{\kappa,\delta}(m,\Phi)=\kappa\,\sqrt{\mu_\delta},
	\qquad \kappa=\pm,\;\;\delta=\pm.
	\label{eq:appC_E_kd_compact}
	\end{equation}
	Expanding \eqref{eq:appC_E_kd_compact} gives the equivalent form used in Sec.~\ref{sec:dirac_ring},
	\begin{equation}
	\begin{aligned}
	E_{\kappa,\delta}
	&=
	\frac{\kappa}{2}\Big[
	\epsilon^2\left(1+4\tilde m^{\,2}\right)+8\lambda_R^2\\
	&\qquad
	-4\delta
	\sqrt{
	\left(\epsilon^2\tilde m^{\,2}+\lambda_R^2\right)
	\left(\epsilon^2+4\lambda_R^2\right)}
	\Big]^{1/2}.
	\end{aligned}
	\label{eq:appC_E_kd_expanded}
	\end{equation}
	
	\subsection{Eigenvectors (closed form)}
	Let $u_\delta$ be a normalized eigenvector of $B_mB_m^\dagger$ with eigenvalue $\mu_\delta$:
	\begin{equation}
	(B_mB_m^\dagger)\,u_\delta=\mu_\delta\,u_\delta,
	\qquad
	u_\delta\in\mathbb{C}^2.
	\end{equation}
	A convenient unnormalized choice obtained from the first row of
	$\left(B_mB_m^\dagger-\mu_\delta\mathbb{I}\right)u_\delta=0$ is
	\begin{equation}
	\tilde u_\delta :=
	\begin{pmatrix}
	2\lambda_R(b-A)\\
	\mu_\delta-(b-A)^2
	\end{pmatrix},
	\qquad
	u_\delta:=\frac{\tilde u_\delta}{\|\tilde u_\delta\|}.
	\label{eq:appC_u_delta}
	\end{equation}
	Now write a four-component eigenvector of $H_m$ as $\psi=(u,v)^{\mathsf T}$ with
	$u,v\in\mathbb{C}^2$ (the $A$- and $B$-pseudospin sectors). The eigenvalue equation
	$H_m\psi=E\psi$ is equivalent to
	\begin{equation}
	B_m v = E u,
	\qquad
	B_m^\dagger u = E v.
	\end{equation}
	For $E_{\kappa,\delta}=\kappa\sqrt{\mu_\delta}\neq 0$ we may choose $u=u_\delta$ and obtain
	\begin{equation}
	v_{\kappa,\delta}=\frac{1}{E_{\kappa,\delta}}\,B_m^\dagger u_\delta
	=\frac{\kappa}{\sqrt{\mu_\delta}}\,B_m^\dagger u_\delta.
	\label{eq:appC_v_from_u}
	\end{equation}
	Thus a normalized eigenvector can be written as
	\begin{equation}
	\psi_{\kappa,\delta}
	=
	\mathcal{N}_{\kappa,\delta}
	\begin{pmatrix}
	u_\delta\\[4pt]
	\frac{\kappa}{\sqrt{\mu_\delta}}\,B_m^\dagger u_\delta
	\end{pmatrix},
	\label{eq:appC_eigenvector}
	\end{equation}
	where
	\begin{equation*}
	\mathcal{N}_{\kappa,\delta}^{-2}
	=
	\|u_\delta\|^2+\frac{1}{\mu_\delta}\|B_m^\dagger u_\delta\|^2
	=2,
	\end{equation*}
	where the final equality follows from $(B_m^\dagger u_\delta)^\dagger(B_m^\dagger u_\delta)
	=u_\delta^\dagger(B_mB_m^\dagger)u_\delta=\mu_\delta\|u_\delta\|^2$.
	Hence $\mathcal{N}_{\kappa,\delta}=1/\sqrt{2}$.
	
	\paragraph{Particle--hole pairing.}
	Because $H_m$ is off-diagonal, the spectrum is symmetric: if $\psi_{+,\delta}$ has energy $+\sqrt{\mu_\delta}$,
	then $\psi_{-,\delta}$ has energy $-\sqrt{\mu_\delta}$ with the same $u_\delta$ and opposite sign in \eqref{eq:appC_v_from_u}.
	
	\subsection{Including intrinsic SOC (optional extension)}
	If intrinsic SOC is included in Sec.~\ref{sec:dirac_ring}, the mode Hamiltonian becomes
	\begin{equation}
	H_m \mapsto H_m + \Delta_{\rm SO}\,\sigma_z\otimes s_z,
	\label{eq:appC_add_intrinsic}
	\end{equation}
	which breaks the strictly off-diagonal pseudospin block structure. The diagonalization is still analytic,
	but the compact singular-value reduction above no longer applies directly; for reproducibility we recommend
	either (i) squaring and reducing to two coupled $2\times 2$ problems using conserved symmetries when present,
	or (ii) diagonalizing the resulting $4\times 4$ matrix numerically and verifying the $\Delta_{\rm SO}\to 0$
	limit recovers \eqref{eq:appC_E_kd_expanded}.
	
	\section{\texorpdfstring{Closed-form $SU(2)$ holonomies in special Rashba--Dresselhaus limits}{Closed-form SU(2) holonomies in special Rashba--Dresselhaus limits}}
	\label{app:RD_limits}
	
	\subsection{Ring holonomy as a Wilson line}
	In the 2DEG ring, the internal transport is encoded by the $SU(2)$ Wilson line
	\begin{equation}
	U_{\SU(2)}
	=
	\mathcal{P}\exp\!\left(-i\int_0^{2\pi} d\varphi\; W_\varphi(\varphi)\right),
	\label{eq:appD_USU2}
	\end{equation}
	where $W_\varphi(\varphi)$ is the tangential component of the effective $SU(2)$ connection determined
	by Rashba ($\alpha$) and Dresselhaus ($\beta$) couplings, as in Sec.~\ref{sec:RD_ring}.
	Path ordering is generally essential because $[W_\varphi(\varphi),W_\varphi(\varphi')]\neq 0$.
	
	Define the dimensionless SOC strengths
	\begin{equation}
	k_R:=\frac{m\alpha a}{\hbar^2},\qquad
	k_D:=\frac{m\beta a}{\hbar^2}.
	\end{equation}
	
	\subsection{\texorpdfstring{Pure Rashba ($\beta=0$): exact removal of $\varphi$-dependence}{Pure Rashba (beta=0): exact removal of phi-dependence}}
	For $\beta=0$ one finds $W_\varphi(\varphi)= -k_R\,\sigma_\rho(\varphi)$, where
	$\sigma_\rho(\varphi)=\sigma_x\cos\varphi+\sigma_y\sin\varphi$.
	Introduce the comoving spin rotation
	\begin{equation}
	\begin{aligned}
	g(\varphi)&=\exp\!\left(-\frac{i}{2}\varphi\,\sigma_z\right),\\
	g^{-1}\sigma_\rho(\varphi)g&=\sigma_x,\\
	i\,g^{-1}\partial_\varphi g&=\frac{1}{2}\sigma_z.
	\end{aligned}
	\end{equation}
	On the interval $[0,2\pi]$, the transformed connection
	\begin{equation}
	W_\varphi\mapsto W'_\varphi=g^{-1}W_\varphi g+i g^{-1}\partial_\varphi g
	\end{equation}
	becomes constant:
	\begin{equation}
	W'_\varphi = -k_R\,\sigma_x + \frac12\sigma_z.
	\label{eq:appD_Wprime_rashba}
	\end{equation}
	Because $g(2\pi)=-\mathbb{I}$ while $g(0)=\mathbb{I}$, this comoving rotation is not periodic on $S^1$.
	Therefore the \emph{closed-loop} holonomy in the original frame is not just the exponential of $W'_\varphi$:
	\begin{equation}
	\begin{aligned}
	U_{\SU(2)}^{(\beta=0)}
	&=
	g(2\pi)\,\exp\!\left(-i\int_0^{2\pi}\!d\varphi\,W'_\varphi\right)\,g(0)^{-1}
	\\
	&=
	-\,\exp\!\Big(-i2\pi\,\bm\Gamma_R\cdot\bm\sigma\Big),
	\end{aligned}
	\label{eq:appD_USU2_rashba}
	\end{equation}
	where
	\begin{equation}
	\bm\Gamma_R=( -k_R,\,0,\,1/2).
	\end{equation}
	Writing $\Gamma_R:=|\bm\Gamma_R|=\sqrt{k_R^2+1/4}$, we obtain
	\begin{equation}
	U_{\SU(2)}^{(\beta=0)}
	=
	\Big[-\cos(2\pi\Gamma_R)\Big]\,\mathbb{I}
	+i\sin(2\pi\Gamma_R)\,\frac{\bm\Gamma_R}{\Gamma_R}\cdot\bm\sigma,
	\label{eq:appD_USU2_rashba_closed}
	\end{equation}
	with eigenvalues $-\exp(\mp i2\pi\Gamma_R)$.
	This passes the consistency check $k_R=0\Rightarrow \Gamma_R=1/2$, for which
	$U_{\SU(2)}^{(\beta=0)}=\mathbb{I}$ as required by the vanishing original connection.
	
	\subsection{\texorpdfstring{Pure Dresselhaus ($\alpha=0$): exact reduction by a shifted comoving frame}{Pure Dresselhaus (alpha=0): exact reduction by a shifted comoving frame}}
	For $\alpha=0$, $W_\varphi(\varphi)$ is proportional to a rotated in-plane Pauli matrix
	$\sigma_\rho(\varphi+\pi/2)$. Defining
	\begin{equation}
	g_D(\varphi)=\exp\!\left(-\frac{i}{2}(\varphi+\pi/2)\sigma_z\right)
	\end{equation}
	one obtains a constant transformed connection of the form
	\begin{equation}
	W'_{\varphi,D}=-k_D\,\sigma_x+\frac12\sigma_z,
	\end{equation}
	with the same endpoint factor $g_D(2\pi)g_D(0)^{-1}=-\mathbb{I}$.
	Therefore the physical loop holonomy is identical to \eqref{eq:appD_USU2_rashba_closed} with $k_R\to k_D$.
	
	\subsection{\texorpdfstring{Pure-gauge locus $\alpha=\pm\beta$: trivial holonomy up to conjugation}{Pure-gauge locus alpha=+-beta: trivial holonomy up to conjugation}}
	In the uniform Rashba--Dresselhaus problem the non-Abelian curvature is commutator-generated and proportional to
	$\alpha^2-\beta^2$, so on the locus $\alpha=\pm\beta$ the $SU(2)$ field strength vanishes and the connection is
	pure gauge. In that case there exists a smooth $g(\bm r)\in SU(2)$ such that $W_i=i\,g^{-1}\partial_i g$,
	and therefore any Wilson line depends only on endpoints:
	\begin{equation}
	U(\gamma)=g(x_f)^{-1}g(x_i).
	\end{equation}
	For a closed loop ($x_f=x_i$) one has $U(C)=\mathbb{I}$ in a simply connected gauge patch (and more generally
	is trivial up to conjugation determined by global/topological data). This is the structural reason the
	ordering hierarchy collapses on $\alpha=\pm\beta$ and provides the null test used in Sec.~\ref{sec:RD_ring}.
	
	\subsection{\texorpdfstring{Leading non-Abelian correction away from $\alpha=\pm\beta$}{Leading non-Abelian correction away from alpha=+-beta}}
	For general $\alpha,\beta$ the noncommutativity of $W_\varphi(\varphi)$ at different angles forces ordering.
	Writing $U_{\SU(2)}=\exp(\Omega_1+\Omega_2+\cdots)$, the leading ordering correction is the Magnus term
	\begin{equation}
	\Omega_2
	=
	-\frac12\int_0^{2\pi}\!d\varphi_1\int_0^{\varphi_1}\!d\varphi_2\;
	\big[W_\varphi(\varphi_1),W_\varphi(\varphi_2)\big],
	\end{equation}
	which vanishes in the commuting/pure-gauge limits and is controlled by the commutator sector
	(i.e.\ by curvature) as developed in Appendix~\ref{app:Magnus}.
	
	\section{Relation to loop and surface representations}
	\label{app:loop_surface_relation}
	
	\subsection{Holonomy algebra viewpoint}
	Our ``second composition'' is naturally phrased in terms of a holonomy (Wilson-line) $\ast$-algebra:
	the configuration data are group elements $U(\gamma)$ assigned to paths $\gamma$, with concatenation mapped to
	multiplication and reversal mapped to inversion. This is precisely the algebraic backbone of the holonomy
	$C^\ast$-algebra approach, where states are positive linear functionals on the completed holonomy algebra and
	the loop transform appears as a representation-theoretic statement. In that setting, ``strip'' (surface-thickened)
	labels emerge naturally alongside loop labels in the spectral theory of the $C^\ast$-algebra.  The condensed-matter
	program here differs only in that $\cA$ is effective/background rather than a dynamical gauge field with Gauss constraints.
	
	\subsection{Reconstruction and completeness (why loop data are enough)}
	A crucial structural fact (used implicitly throughout) is that Wilson-line/loop data can encode the gauge potential
	(up to gauge) under appropriate regularity assumptions: open-path phases and loop variables satisfy algebraic relations
	sufficient for reconstruction. This is one reason holonomy is the ``right'' geometric variable when the physics is
	interferometric and topology-sensitive: it packages gauge covariance and path composition in a single object.
	
	\subsection{Dual loop/path representations and multivaluedness}
	In dual/generalized loop representations of Abelian gauge theories with sources or topological defects, wave functionals
	often become multivalued and acquire dependence on spanning surfaces bounded by loops. This phenomenon is the geometric
	counterpart of how multiply connected configuration spaces produce topological phases in ordinary quantum mechanics.
	In the loop-representation literature this is made explicit for Maxwell theory with monopoles or charges: the nonlocal
	operators (Wilson-like and disorder/'t~Hooft-like duals) form a topological algebra, and surface dependence disappears
	precisely when appropriate quantization conditions hold. Conceptually, this parallels our loop--surface lift:
	surface variables (curvature fluxes, Stokes data) are not optional decorations but the natural carriers of topological
	and ordering information.
	
	\subsection{Positioning of the present work}
	With these viewpoints in mind, the present manuscript can be read as follows:
	\begin{itemize}
	\item The first composition (Sec.~\ref{sec:covariant_structure}) identifies an effective $\Uone$ connection together with the relevant internal non-Abelian connection
	      from quantum-matter Hamiltonians (Pauli and Dirac).
	\item The second composition (Sec.~\ref{sec:loop_algebra}) places the theory in a holonomy/loop algebra, preserving
	      the $\ast$-algebraic structure and making topology explicit through $\pi_1(M)$.
	\item The non-Abelian Stokes theorem (Appendix~\ref{app:NAST}) and Magnus hierarchy (Appendix~\ref{app:Magnus})
	      provide the surface and commutator calculi that, in loop/surface representations, encode the same structural
	      content: holonomy lives on loops; non-Abelianity lives in ordering/curvature; topology is carried by how loops
	      bound (or fail to bound) surfaces.
	\end{itemize}
	In this sense, our program is a condensed-matter realization of loop/surface logic, but directed at transport and
	interference rather than at quantization of a dynamical gauge field.
	
	\section{Explicit Rashba--Dresselhaus gauge potentials and curvature}
	\label{app:RD_potentials_curvature}
	
	\subsection{\texorpdfstring{Consistent convention for $\alpha,\beta$ and the $SU(2)$ gauge field}{Consistent convention for alpha, beta and the SU(2) gauge field}}
	To match the dimensionless parameters
	$k_R=m\alpha a/\hbar^2$ and $k_D=m\beta a/\hbar^2$ used in Appendix~\ref{app:RD_limits},
	it is convenient to write the SOC Hamiltonian in the standard form
	\begin{equation}
	H_{\rm SO}=
	\frac{\alpha}{\hbar}\big(\sigma_x p_y-\sigma_y p_x\big)
	+\frac{\beta}{\hbar}\big(\sigma_x p_x-\sigma_y p_y\big),
	\label{eq:appF_HSO_convention}
	\end{equation}
	so that $\alpha,\beta$ have units of energy$\times$length.
	
	Then the minimal-coupling form
	\begin{equation}
	H=\frac{1}{2m}\big(\bm p+e\bm A-\bm{\mathcal A}\big)^2
	+V_{\rm conf}-\frac{1}{2m}\bm{\mathcal A}^{\,2}
	\label{eq:appF_min_coupling}
	\end{equation}
	reproduces \eqref{eq:appF_HSO_convention} with
	\begin{equation}
	\mathcal A_x=\frac{m}{\hbar}\big(\alpha\,\sigma_y-\beta\,\sigma_x\big),\qquad
	\mathcal A_y=\frac{m}{\hbar}\big(\beta\,\sigma_y-\alpha\,\sigma_x\big),
	\label{eq:appF_Axy}
	\end{equation}
	and hence the dimensionless $SU(2)$ connection (used throughout the main text)
	\begin{equation}
	\begin{aligned}
	W_i&\equiv \frac{1}{\hbar}\mathcal A_i,\\
	W_x&=\frac{m}{\hbar^2}\big(\alpha\,\sigma_y-\beta\,\sigma_x\big),\\
	W_y&=\frac{m}{\hbar^2}\big(\beta\,\sigma_y-\alpha\,\sigma_x\big).
	\end{aligned}
	\label{eq:appF_Wxy}
	\end{equation}
	This is the standard ``SOC as $SU(2)$ gauge field'' identification used in ring interferometry formulations.
	\cite{Hatano2007}
	
	\subsection{Tangential connection on the ring}
	On a ring of radius $a$ in the $xy$-plane, the unit tangent is
	$\hat{\bm e}_\varphi=(-\sin\varphi,\cos\varphi)$, so the tangential connection is
	\begin{equation}
	W_\varphi(\varphi)=a\,\hat{\bm e}_\varphi\!\cdot\!\bm W
	=a\big(-\sin\varphi\,W_x+\cos\varphi\,W_y\big).
	\end{equation}
	Substituting \eqref{eq:appF_Wxy} gives the explicit rotating-axis form
	\begin{equation}
	\begin{split}
	W_\varphi(\varphi)
	&=\frac{ma}{\hbar^2}\Big[
	\big(\beta\cos\varphi-\alpha\sin\varphi\big)\sigma_y\\
	&\qquad+\big(\beta\sin\varphi-\alpha\cos\varphi\big)\sigma_x\Big].
	\end{split}
	\label{eq:appF_Wphi_explicit}
	\end{equation}
	For generic $(\alpha,\beta)$ the spin-space axis rotates with $\varphi$, implying
	$[W_\varphi(\varphi),W_\varphi(\varphi')]\neq 0$ and necessitating path ordering in the Wilson line.
	
	\subsection{\texorpdfstring{$SU(2)$ curvature in the uniform-coupling case}{SU(2) curvature in the uniform-coupling case}}
	For uniform $\alpha,\beta$ the derivative terms vanish and the curvature is commutator-generated:
	\begin{equation}
	F_{xy}=\partial_x W_y-\partial_y W_x-i[W_x,W_y]= -i[W_x,W_y].
	\end{equation}
	Using \eqref{eq:appF_Wxy} and $[\sigma_x,\sigma_y]=2i\sigma_z$ yields
	\begin{equation}
	F_{xy}
	= +2\left(\frac{m}{\hbar^2}\right)^2(\alpha^2-\beta^2)\,\sigma_z,
	\label{eq:appF_Fxy}
	\end{equation}
	consistent with \eqref{eq:appF_Wxy} and the identity $[\sigma_x,\sigma_y]=2i\sigma_z$.
	This agrees with \eqref{eq:Fxy_RD} in the main text.
	The structural content is invariant under sign redefinitions:
	\begin{equation}
	F_{xy}=0\quad\Longleftrightarrow\quad \alpha=\pm\beta,
	\end{equation}
	which is the pure-gauge checkpoint emphasized in the main text and in the spin-helix literature.
	\cite{TokatlySherman2009}
	
	\subsection{\texorpdfstring{Explicit pure-gauge form and the gauging-away transformation at $\alpha=\pm\beta$}{Explicit pure-gauge form and the gauging-away transformation at alpha=+-beta}}
	When $\alpha=\pm\beta$, one finds that $W_x$ and $W_y$ become proportional (hence commute), so the connection is
	pure gauge: there exists a smooth $g(\bm r)\in SU(2)$ such that
	\begin{equation}
	W_i(\bm r)= i\,(\partial_i g)g^{-1}.
	\label{eq:appF_puregauge_condition}
	\end{equation}
	A constructive choice (valid when $[W_x,W_y]=0$) is
	\begin{equation}
	\begin{aligned}
	g(x,y)&=\exp\!\big(-i(xW_x+yW_y)\big),\\
	\Rightarrow\qquad
	i(\partial_i g)g^{-1}&=W_i.
	\end{aligned}
	\label{eq:appF_g_choice}
	\end{equation}
	Then the gauge transformation by $g^{-1}$ removes the connection:
	\begin{equation}
	W_i \mapsto W_i^{\,g^{-1}} = g^{-1}W_i g + i(\partial_i g^{-1})g =0.
	\end{equation}
	This is precisely the mechanism behind the ``pure gauge'' regime in which equilibrium spin currents vanish
	and conserved spin projections emerge, as emphasized by Tokatly and Sherman.
	\cite{TokatlySherman2009}
	
	\subsection{Implication for Wilson loops on the ring}
	If $W_i$ is pure gauge, any Wilson line depends only on endpoints:
	\begin{equation}
	U(\gamma)=\mathcal{P}\exp\!\left(-i\int_\gamma W\right)=g(x_f)^{-1}g(x_i),
	\end{equation}
	so for a closed loop ($x_f=x_i$) the $SU(2)$ holonomy is trivial (or, more precisely, trivial up to a global
	conjugation fixed by patching/topology). Therefore on $\alpha=\pm\beta$ the internal contribution to the
	ring holonomy carries no intrinsic non-Abelian content, matching the null-test logic used in
	Sec.~\ref{sec:RD_ring}.
	
	\subsection{Connection to the persistent-spin-helix symmetry point}
	The same $\alpha=\pm\beta$ condition underlies the exact $SU(2)$ symmetry responsible for persistent spin helix
	behavior in Rashba--Dresselhaus systems. \cite{BernevigOrensteinZhang2006,Koralek2009}
	The experimental realization of this helix in GaAs quantum wells \cite{Koralek2009} confirms that the pure-gauge
	condition $\alpha=\pm\beta$ is not merely a formal checkpoint but a physically accessible regime.
	In the gauge-field language, that symmetry point is exactly the vanishing-curvature (pure-gauge) locus.

\end{appendix}

\section*{Acknowledgements}
The author gratefully acknowledges the support of the Escuela de F\'isica, Universidad Central de
Venezuela, and of Astrum Drive Technologies, where this work was carried out.

\paragraph{Funding information}
This research received no specific grant from any funding agency in the public, commercial, or
not-for-profit sectors. N.B.\ acknowledges institutional support from the Universidad Central de
Venezuela and from Astrum Drive Technologies.

\bibliography{refs}

\nolinenumbers

\end{document}